\documentclass[
  journal=pasa,
  manuscript=research-paper, 
  year=2024,
  volume=00,
]{cup-journal}

\usepackage{orcidlink}
\usepackage{cuted2}
\usepackage{microtype,siunitx,booktabs}
\usepackage{CJKutf8}
\usepackage{xcolor}
\usepackage[normalem]{ulem}

\usepackage{tikz}

\usepackage{longtable}
\usepackage{graphicx}
\usepackage{amsmath}
\usepackage{amssymb}
\usepackage{booktabs}
\usepackage{subcaption}
\usepackage{makecell}
\usepackage{xurl}
\usepackage{hyperref}
\definecolor{dodgerblue}{RGB}{30, 144, 255}
\definecolor{crimson}{RGB}{220, 20, 60}
\definecolor{darkerblue}{RGB}{0, 0, 139}
\hypersetup{colorlinks,citecolor=dodgerblue,linkcolor=blue,urlcolor=darkerblue}

\usepackage{siunitx}

\usepackage[nolist,nohyperlinks]{acronym}

\usepackage{listings}
\DeclareSIUnit\degreesq{deg\text{$^2$}}
\DeclareSIUnit\perdegreesq{deg\text{$^{-2}$}}

\graphicspath{{./}{figures/}}

\title{How to Count Clustered Galaxies}

\author{Yunting Wang \begin{CJK}{UTF8}{gbsn}
(王允婷)\end{CJK} \orcidlink{0000-0001-5751-2347}}
\affiliation{Department of Physics and Astronomy, University of British Columbia, 6225 Agricultural Road, Vancouver V6T 1Z1, Canada}
\email[Yunting Wang]{yuntingw@phas.ubc.ca}

\author{Ryley Hill}
\affiliation{Department of Physics and Astronomy, University of British Columbia, 6225 Agricultural Road, Vancouver V6T 1Z1, Canada}

\author{Douglas Scott\orcidlink{0000-0002-6878-9840}}
\affiliation{Department of Physics and Astronomy, University of British Columbia, 6225 Agricultural Road, Vancouver V6T 1Z1, Canada}

\author{Tessa Vernstrom}
\affiliation{Australia Telescope National Facility, CSIRO Space and Astronomy, PO Box 1130, Bentley WA 6102, Australia}
\alsoaffiliation{ICRAR, The University of Western Australia, 35 Stirling Hw, 6009 Crawley, Australia}

\doi{00.0000/pasa.0000.00}

\received {dd Mmm YYYY}
\revised  {dd Mmm YYYY}
\accepted {dd Mmm YYYY}
\published{00 Mmm YYY}


\begin{document}

\begin{abstract}
Obtaining robust galaxy number counts is crucial for understanding galaxy evolution, and submillimetre counts in particular have proven valuable for revising subgrid physics models in cosmological simulations. In confusion-limited surveys, which are common at these wavelengths, statistical methods such as $P(D)$ fluctuation analysis are required to recover counts of faint, unresolved galaxies. 
However, the standard $P(D)$ framework assumes that galaxies are Poisson-distributed, whereas in reality galaxies are clustered. Using simulations, we demonstrate that this clustering systematically biases $P(D)$-derived number counts, and present an empirical method that simultaneously measures and corrects for this bias by combining the 1- and 2-point statistics in the map, thereby maximising the information extracted from the data. Applying this method to deep {\it Herschel}-SPIRE observations of the GOODS-N field, we provide revised galaxy number counts at 250, 350 and 500\,$\mu$m. Our results indicate that at 500\,$\mu$m clustering inflates the apparent counts by a factor of 1.6 around $10\,$mJy and slightly suppresses the faintest sub-mJy counts, with milder effects at 350\,$\mu$m and 250\,$\mu$m owing to the smaller beam sizes. This methodology is broadly applicable to other confusion-limited data sets with well-characterised beam and noise properties, including {\it Spitzer}, SCUBA-2, and the upcoming CCAT, enabling unbiased exploitation of the full statistical information in many surveys in the mid-IR to radio wavebands.

\end{abstract}

\section{Introduction }\label{sec:intro}

Galaxy number counts, the number density of galaxies as a function of flux density, are a powerful statistical tool for understanding galaxy evolution. They represent the observed projection of the luminosity function in the absence of redshift information, providing important constraints on subgrid physics in cosmological simulations. The submillimetre wavelength regime is particularly sensitive to dust-obscured star formation and has historically been effective at selecting high-redshift galaxies over large cosmic volumes, yet the observed submillimetre number counts remain difficult to reproduce with optically-tuned simulations, pointing to the need for revised feedback prescriptions \citep[e.g.][]{Lacey2016,Lovell2021,Parmar2026}.

Despite their scientific merit, measurements at the faint end of the submillimetre number counts are often confusion limited. At these wavelengths, telescope beams are large, and faint galaxies below the detection threshold cannot be individually resolved, but instead contribute collectively to background fluctuations, known as confusion noise. Conventional source-extraction methods are restricted to fluxes above the confusion limit, while high-resolution surveys that circumvent this limitation typically cover only small fields of view.

Unlike instrumental noise, confusion noise carries genuine astrophysical information about the underlying source population near and below the confusion limit, which can be analysed with statistical methods. Two techniques are commonly used to probe below this threshold: stacking and $P(D)$ fluctuation analysis. In stacking, the positions of galaxies identified through deep observations at other wavelengths, such as the near-infrared, are used as priors to stack fluxes at the target wavelength \citep[e.g.][]{Marsden2009}. If redshift information is known from the catalogue, stacking can be performed in different redshift slices to obtain galaxy number counts at different redshift bins \citep[e.g.][]{Bethermin2012, Viero2013, Duivenvoorden2020, Hill2025}.

The second approach, $P(D)$, is the focus of this paper. $P(D)$ fluctuation analysis is a statistical framework that characterises the contribution from faint, unresolved sources to the 1-point statistics of the map, i.e., the observed pixel flux histogram. While direct counting effectively takes the sources contributing to the extreme positive (bright) tail of the histogram, $P(D)$ leverages information from the entire shape of the flux distribution. In comparison to stacking, $P(D)$ analysis recovers statistical information from the map alone, requiring no prior knowledge from other wavelengths and thereby avoiding observational biases introduced by incomplete prior catalogues. The name $P(D)$ derives from the historical term `deflection', referring to the motion of a chart recorder needle \citep{Scheuer1957}, where $D$ now denotes pixel intensity in units of flux density per beam solid angle.

The $P(D)$ distribution represents the brightness function of the underlying population convolved with the telescope beam and broadened by the instrumental noise, and is therefore characterised by both the beam and noise properties. By minimising the difference between the predicted $P(D)$ for input galaxy count models and the observed flux histograms through a Markov chain Monte Carlo (MCMC) approach, $P(D)$ has successfully yielded deep galaxy number counts from far-infrared (FIR) to radio wavelengths \citep[e.g.][]{Patanchon2009, Glenn2010, Bethermin2012, Vernstrom2014, Wang2017, Mauch2020}. Specifically for \textit{Herschel}-Spectral and Photometric Imaging Receiver (SPIRE; \citealt{Griffin2010}), \citet{Glenn2010} applied $P(D)$ analysis on early \textit{Herschel} Multi-tiered Extragalactic Survey \citep[HerMES;][]{Oliver2010} data, while \citet{Bethermin2012} later conducted a stacking analysis on updated SPIRE maps. However, a $P(D)$ analysis on the final, highest signal-to-noise SPIRE maps remains absent from the established research.

A fundamental limitation of standard $P(D)$ analysis, however, is the assumption that galaxies are Poisson-distributed in their positions on the sky. In reality, galaxies are clustered, which can bias the galaxy number counts derived through the $P(D)$ framework \citep[see e.g.][]{Bethermin2017}. Studies incorporating source clustering in $P(D)$ analyses have been explored analytically \citep[e.g.][]{Barcons1992, Toffolatti1998}. In particular \citet{Takeuchi2004} presented an analytical formulation of confusion noise and source clustering using the assumption that galaxies are clustered with a uniform clustering strength. The formulation is applicable to simple analytical galaxy counts models, such as the power-law model, but introduces increasing complexity in modelling realistic galaxy count models. 

Empirical characterisation of clustering in confused maps has been explored only recently with the Simulated Infrared Dusty Extragalactic Sky (SIDES; \citealt{Bethermin2017}) simulation, which contains realistic clustering from dark matter halo simulations. \citet{Bethermin2017} demonstrated that clustering can distort the flux histogram of maps, especially for large telescope beams, thus biasing the $P(D)$ galaxy counts fit to the histogram. In one approach to address this problem, \citet{Bing2023} recently introduced a correction for clustering bias by estimating the ratio between the true SIDES counts and the direct counts extracted on the simulated (clustered) SIDES maps. However, this method assumes that the observed field strictly mirrors the counts and clustering properties of the SIDES simulation. It is clear that a self-consistent, data-driven approach would be preferable.

For images of unresolved galaxies, it is expected that the vast majority of information is contained within the 1-point and 2-point functions; higher-order statistical information is present, but its signal tends to be weak. It therefore makes sense to estimate the 1- and 2-point functions simultaneously, and correct how one influences the other.
The clustering information of galaxies is encoded in the 2-point correlation function (2PCF) or its Fourier-space analogue, the angular power spectrum. In this paper, we present an empirical framework that combines 2-point statistics with $P(D)$ analysis to effectively correct galaxy number counts for clustering bias. Motivated by the wealth of existing and upcoming FIR-to-radio surveys, this methodology provides a robust, generalised tool for extracting more accurate intrinsic counts from confused maps.

The structure of this paper is as follows. Section~\ref{ch:pd} describes the basic approach of the $P(D)$ analysis. In Section~\ref{ch:Models}, we test the significance of clustering bias in galaxy number counts using submillimetre simulations, demonstrating the non-negligible effect of clustering for large telescope beams. Section~\ref{ch:sim} systematically characterises the impact of clustering on the histogram by simulating maps of different clustering strengths. We then develop and validate a correction method that directly combines the 1- and 2-point statistical information from a map. In Section~\ref{ch:goodsn}, we apply this correction method to full-depth \textit{Herschel}-SPIRE data of the GOODS-N field at 250, 350, and 500\,$\mu$m. Clustering leads to an overestimate of the 500-$\mu$m counts by a factor of 1.6 at around 10\,mJy, with smaller effects for the other SPIRE wavebands. In Section~\ref{ch:discussion} we show that these systematic biases of clustering can be corrected by estimating the strength of the clustering, and we provide revised counts at 250, 350, and 500\,$\mu$m.  We end with our conclusions in Section~\ref{ch:conclusions}.

\section{\textit{P(D)} fluctuation analysis}
\label{ch:pd}

$P(D)$ effectively quantifies the process of the collective observed response of an intrinsic distribution of flux signals. The signal response of a point source with flux density \textit{S} for a beam function \textit{B} is $x=S B(\theta, \phi)$.\footnote{The beam function $B(\theta, \phi)$, also called the point spread function or PSF, is peak-normalised here, to be consistent with $x$ in units of flux density per beam solid angle.} The intrinsic integral number count can be written $N(>S)$, which is the total number of sources per beam solid angle with flux density larger than \textit{S}, while the differential number counts per beam solid angle is then $dN/dS$. Therefore, the mean number of sources per beam solid angle with observed flux between $x$ and $x+dx$ is
\begin{equation}
R(x)dx = \int_{\Omega} \frac{d N}{d S} d \Omega d S=\int_{\Omega} \frac{d N}{d S} B(\theta, \phi)^{-1} d \Omega d x.
\label{eq:Rx}
\end{equation}
Let us assume that the number of sources with observed flux density \textit{x} follows a Poisson distribution, i.e.~a random spatial distribution, with the mean number given in equation (\ref{eq:Rx}). Define the sum of flux densities from all sources as $D$ (the observed pixel values).
The probability density function (PDF) of $D$, $P(D)$, can be derived by multiplying the individual characteristic functions in the Fourier plane (see \citealt{Condon1974}, \citealt{Wall1982}, \citealt{Takeuchi2001}, \citealt{Patanchon2009} for detailed derivations),
\begin{equation}
\begin{aligned}
P(D)&=\mathcal{F}^{-1}\Biggl[\exp \Biggl(\int_{0}^{\infty} R(x) \exp (i \omega x) d x\\ &-\int_{0}^{\infty} R(x) d x\Biggl)\Biggl],
\end{aligned}
\label{eq:pd}
\end{equation}
where $\omega$ is the variable in Fourier space. The noise in the data is assumed to follow a Gaussian distribution with standard deviation $\sigma_n$, which adds a multiplication term in the Fourier plane and $P(D)$ becomes
\begin{equation}
\begin{aligned}
P(D)&=\mathcal{F}^{-1}\Biggl[\exp \Biggl(\int_{0}^{\infty} R(x) \exp (i \omega x) d x\\ &-\int_{0}^{\infty} R(x) d x - \frac{\sigma_{n}^{2} \omega^{2}}{2}\Biggl) \Biggl].
\end{aligned}
\label{eq:pd_n}
\end{equation}

The PDF of $D$, the observed flux density in a pixel, is simply a continuous version of the normalised histogram of the map. The effect of spatial clustering of galaxies, which is not included in the $P(D)$ framework, is rather small in many applications. However, \citet{Bethermin2017} found in simulations that clustering biases the flux histograms of maps from \textit{Herschel}-SPIRE, and the effect is most significant at 500\,$\mu$m, where the beam size is the largest. This makes sense, since sources clustered within a beam could be interpreted as a single brighter source, thereby changing the pixel distribution. In the next section we will use simulations of clustered and unclustered sources to build a method to correct for the effects of clustering.

\section{Model fits}
\label{ch:Models}

In this section, we first justify the model used throughout by simulating SPIRE observations using the SIDES simulation (Section~\ref{sec:simulation-maps}).  We then
introduce the node-based models used in fitting the galaxy number counts and describe how the model parameters are chosen (Section~\ref{ch:Setup}).  We note in Section~\ref{sec:pixel} that the effect of pixel correlations is non-negligible and describe how to treat it properly. In Section~\ref{sec:pdfit}, we implement the $P(D)$ model fit on simulated maps, with galaxies randomly distributed and clustered on the sky following the SIDES simulation. We demonstrate that the $P(D)$ counts fit is indeed biased by clustering.

\subsection{Simulated maps}
\label{sec:simulation-maps}

In preparation for testing the model set-up, we simulate SPIRE observations using the full galaxy catalogues from SIDES. The SIDES simulation is a $2 \, \mathrm{deg}^2$ mock galaxy catalogue at FIR to submillimetre wavelengths over the redshift range $0<z<7$ . Galaxies in SIDES are clustered through abundance matching of a dark matter simulation. While more recent simulations incorporating gas dynamics and dust modelling are now available, SIDES remains well-suited for our purposes, since the key requirement here is realistic clustering from dark matter halos rather than detailed baryonic physics. Figure~12 in \citet{Bethermin2017} illustrates this clustering effect in SIDES, showing the difference between the pixel histogram of the map with clustered galaxy positions and one with galaxy positions randomly distributed.
In that paper it is thus proposed that spatial clustering of galaxies can bias the $P(D)$ results of galaxy counts. The effect should be most significant in the SPIRE 500-$\mathrm{\mu m}$ observations, given the larger beam at the longer wavelengths. 

To illustrate the direct impact on the counts fit from clustering, we generate simulated \textit{Herschel}-SPIRE 500-$\mathrm{\mu m}$ maps in a similar manner to \citet{Bethermin2017}. We obtain the full galaxy catalogues with simulated \textit{Herschel}-SPIRE 500-$\mathrm{\mu m}$ flux densities from SIDES to generate the clustered map. We then randomise the positions in the catalogue and generate another map, hereafter called a `randomised map'. 
We adopt a pixel size of 12$\, \mathrm{arcsec}$ in the SPIRE 500-$\mathrm{\mu m}$ maps. Both maps are convolved with a Gaussian beam of 36.3$\,$arcsec FWHM, which corresponds to the SPIRE beam at 500\,$\mathrm{\mu m}$. We check that the confusion noise ($6.8$\,mJy at this step) is consistent with the measurements of \citet{Nguyen2010}. We then assume Gaussian instrumental noise with $\sigma = 1.03\,  \mathrm{mJy}\, \mathrm{beam^{-1}}$, \footnote{The unit $\mathrm{Jy}\, \mathrm{beam}^{-1}$ is commonly used at submillimetre wavelengths, with the Gaussian convolution kernel peak-normalised. This is in contrast to $\mathrm{Jy}$ units, which gives the flux density of a source without considering the observational effect of the beam.} which is the typical noise value in the central region of the deep GOODS-N field mapped by \textit{Herschel}-SPIRE at 500$\,\mu$m (see Section~\ref{ch:goodsn} for details about GOODS-N). We finally subtract the mean, since \textit{Herschel} is not sensitive to the absolute value in its observations. The image edges are slightly trimmed by 5 pixels to avoid edge effects in the convolution step. 

\subsection{Model parameterisation}
\label{ch:Setup}
To enable Bayesian inference through $P(D)$ to fit underlying galaxy number counts, we parameterise the galaxy number counts using a node-based model with interpolation of multiple power laws; in other words we have a function that is composed of straight-line segments on a log-log
diagram. See \citet{Vernstrom2014} for a discussion of the advantages of this approach compared with other parameterisations.

The positions of the nodes are selected to smoothly model the galaxy number counts. Number counts outside the node range are assumed to be zero. As a result, the faint limit of the nodes is particularly important and needs to be treated on a case-by-case basis, depending on the beam and noise of the data of interest. The first node is often treated as an upper limit in previous studies, since at this flux-density level galaxies do not change the shape of $P(D)$ anymore, but only move the mean of $P(D)$. While the mean can be calculated in the model set-up stage, we note later in this work that the mean (or equivalently the peak position of the histogram) should be fit in order to properly apply the clustering correction (see Section~\ref{sec: clustered}), and hence the mean is not computed in our model and the first node remains an upper limit.

In practice, we choose the first node using the full galaxy catalogue in the simulation and simulate maps of the beam and the noise level for SPIRE (see Section~\ref{sec:pdfit}). We then calculate $P(D)$ with the catalogue counts sampled in nodes and compare $P(D)$ with the histogram of the simulated maps. We determine the first node where the map histogram is consistent with the $P(D)$ calculated from the truncated counts, including the mean of $P(D)$, i.e.\ the data are sensitive down to this flux-density limit. The second node is chosen to be at around a quarter of the instrumental noise level, following \citet{Vernstrom2014}. The brightest node is at around $0.1$\,Jy, above which the number counts become well constrained by directly counting the resolved sources in SPIRE maps \citep[e.g.][]{Oliver2010, Clements2010, Bethermin2012, Valiante2016}. The nodes between the second node and the brightest node are spaced logarithmically. The number of nodes is a trade-off between a full characterisation of the galaxy number counts model shape and computational speed, with too many nodes yielding highly-correlated results. After computing tests on different numbers of nodes on simulated SPIRE maps (see Section ~\ref{sec:pdfit}), we find that eight or nine nodes are sufficient to characterise the counts in SPIRE maps.

We apply some conditions on the model to ensure a physical parameterisation, as follows. (1) The integrated source count flux density $\int S\frac{dN}{dS} dS$ should be consistent with the upper limit of the total estimated cosmic infrared background (CIB) at 250, 350 and 500$\,\mathrm{\mu m}$ from \citet{Fixsen1998}. (2) The slope should not be steeper than $\mathrm{-6}$ between adjacent nodes on a log-log scale. (3) The difference between two adjacent slopes should not be greater than 3.5, to maintain relatively smooth interpolation. (4) The magnitude of the faintest node should be between $10^9$ and $10^{16}$\, Jy$^{-1}$sr$^{-1}$, a weak constraint based on observed galaxy number counts at these wavelengths \citep{Marsden2009, Glenn2010, Bethermin2012}.

We use the MCMC ensemble sampler \texttt{emcee} \citep{Goodman2010, ForemanMackey2013} to fit the multinode model. For each step in the MCMC sampling, the model is evaluated by generating its $P(D)$ prediction in Equation~\ref{eq:pd_n}, and then comparing with the histogram of the input map to define the likelihood. We choose to estimate the log-likelihood as in \citet{Vernstrom2014}:

\begin{equation}
\centering
\log \mathcal{L}=-\frac{1}{2} \sum_{i} \frac{\left(D_{i}-M_{i}\right)^{2}}{\sigma_{i}^{2}},
\label{eq:chi}
\end{equation}

\noindent where $D_i$ is the intensity in the ith bin, 
$\sigma_i$ is the error on $D_i$, and $M_i$ is the $P(D)$-modelled histogram in the $i$th pixel for a given counts model. We assume the uncertainty to be Poisson distributed. We discuss reducing the uncertainty from pixel correlations tested on simulated maps in Section~\ref{sec:pixel}, and from the discretisation process of binning in Section\,\ref{sec:pdfit}.

\subsection{Pixel correlations}
\label{sec:pixel}

\begin{figure}[htbp!]
\centering
	\includegraphics[width=1.0\textwidth]{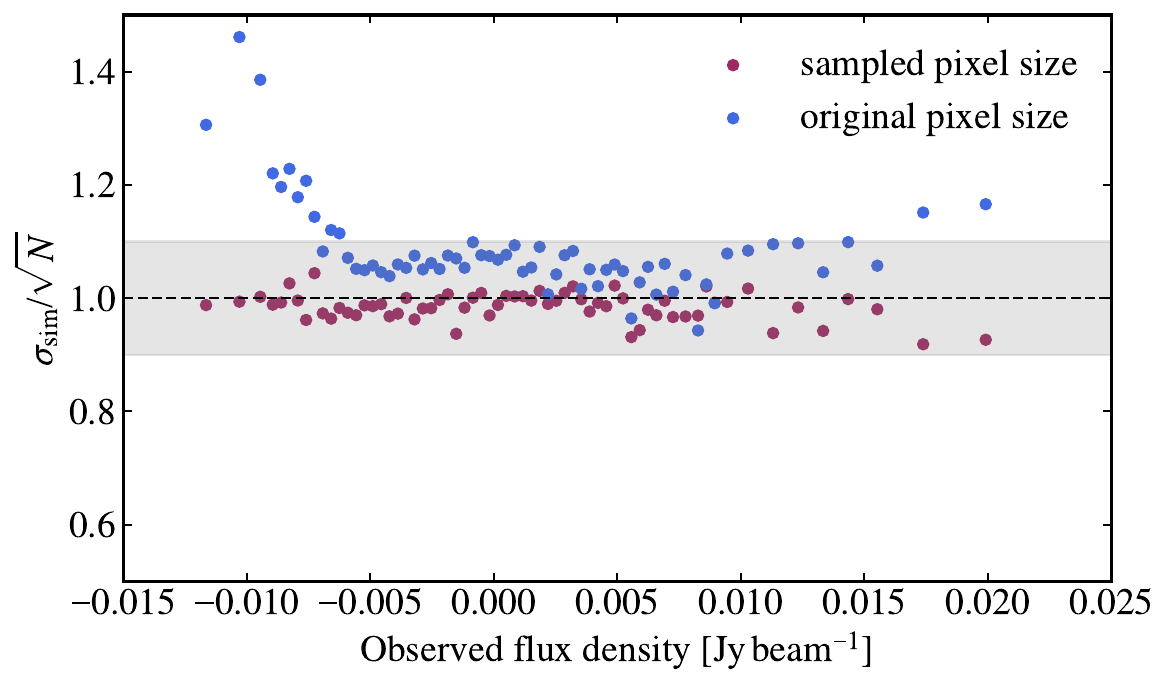}
    \caption{Standard deviation by bin in the histograms of 1000 randomised maps as a ratio of the Poisson error estimate. Randomised maps are simulated at the original pixel size and also with pixels sampled at $1\times$FWHM. A grey band spanning 10\% of the Poisson error is illustrated for visualisation. The sampled map is better represented by Poisson error, while the original map has additional scatter, mostly from pixel correlations.}
    \label{fig:pix}
\end{figure}

The likelihood we adopt in Equation~\ref{eq:chi} assumes Poisson errors. However, the SPIRE beams are larger than the pixel area, where the pixel width is roughly a third of the FWHM of the beam by default in map making \citep{Oliver2012}. The unresolved sources can therefore contribute to multiple pixels, generating correlations between pixels. Ideally, the error in the likelihood should be evaluated using the full covariance matrix of pixel intensities. However, the exact form of the covariance matrix is challenging to determine in real data. Although it is feasible to obtain the covariance matrix through simulations \citep[e.g.][]{Glenn2010}, the underlying counts model adopted in the simulation will have an effect \citep{Vernstrom2014}, as well as non-idealities in the data. Following \citet{Vernstrom2014}, we avoid introducing potential biases from the covariance matrix calculated from the simulations, and instead test the performance of sampling maps at larger pixel sizes. 

We first explore the effect of pixel correlations contributing to the histogram bin error in simulations using the full galaxy catalogues from SIDES at 500\,$\mu$m, where the beam is the largest. We use the 1000 simulated maps with positions of galaxies randomised, as discussed in Section~\ref{sec:simulation-maps}, with the pixel width the same as in the observed 500-$\mu$m map (12 arcsec$\,\simeq\,$FWHM/3), and obtain an average histogram using the binning mentioned in Section~\ref{sec:pdfit}.  The error in each histogram bin can be described by the standard deviation of the number of pixels in the bin in 1000 simulated maps, $\sigma_\text{sim}$. On the other hand, we can calculate the Poisson error by taking the square-root of the number of pixels in the $i$th bin, $\sigma_\text{Poisson} = \sqrt{N_i}$. A comparison between the two should tell us how well the error is described by the Poisson distribution. We repeat the same process for simulated maps sampled at different pixel sizes. In Figure~\ref{fig:pix}, we specifically plot $\sigma_\text{sim}/\sigma_\text{Poisson}$ in each bin for the original pixel size (FWHM/3) and for a pixel size equal to the FWHM. The reference line is $\sigma_\text{sim}/\sigma_\text{Poisson}=1$, where the Poisson error characterises well the error seen in the simulations. In the case of the original pixel width, we find that the overall simulated error is larger than the Poisson error, suggesting that there are underlying correlations contributing on top of the Poisson errors, especially at the bright and faint ends, while the larger sampled pixel size behaves consistently around $\sigma_\text{sim}/\sigma_\text{Poisson} \simeq 1$. We therefore use $1\times$FWHM as the pixel size throughout the work, for all three frequencies of \textit{Herschel}-SPIRE data.

\subsection{\textit{P(D)} fitting}
\label{sec:pdfit}

We implement the $P(D)$ fit of the node-based model (Section~\ref{ch:Setup}) on the histograms of both the clustered and randomised maps.
The numerical integration of the $P(D)$ prediction in Equation~\ref{eq:pd} is calculated in $2^{18}$ flux density bins in the map flux range. In Equation~\ref{eq:chi}, the likelihood is evaluated at each histogram bin, so that the choice of bins is important to trace the statistical significance in the comparison between data and model. We use the simulated clustered map as a reference to create the binning. The map is first sampled at the FWHM pixel width to reduce pixel correlations (Section~\ref{sec:pixel}). We make extra fine ($2^{10}$ to $2^{12}$) linearly uniform bins across the flux density range of the sampled map. We then adaptively merge the fine bins to ensure more than 20 pixels in each bin for Poisson statistics to be reliable. When the number of pixels in a bin is smaller than a fraction of the number of pixels in the most populated bin (peak fraction $1/70$ to $1/50$), we switch to log-spaced bins. In practice we cut off the histogram where the adaptive bins become log-spaced, since we find that wider bins are not well characterised by their central fluxes and may bias the likelihood. 

To ensure our results are robust to the choice of histogram binning, we perform a convergence test on the bin resolution. We vary the adaptive binning parameters (increasing the resolution of the initial linear grid and reducing the peak-fraction threshold) while maintaining the requirement of $N>20$ counts per bin to preserve Gaussian statistics.
We found that beyond a certain resolution, the inferred model parameters and uncertainties stabilised and became independent of the binning scheme. We adopt the finer binning configuration as the optimal setup for our analysis to minimise any discretisation errors, while ensuring the binning is limited only by the map noise.
We note that this adaptive binning is also treated case-by-case. In the following sections when we deal with maps of different size, beam, or noise, we rerun the adaptive binning process to find the optimal binning to reliably evaluate the model likelihood.
Finally, we fix the brightest node to the simulation value, which is obtained through binning the simulation catalogue into fine flux-density bins and interpolating at that node.

Since the maps are not sensitive to the mean brightness in the field, the observed histograms are shifted compared to the $P(D)$ predictions. We simply calculate the mean in each $P(D)$ prediction and shift the $P(D)$ histogram by this mean. We will show later that the clustering bias effectively changes the mean flux in the map, therefore fixing the mean flux to the value predicted by $P(D)$ would not be desirable for clustering corrections (Section~\ref{sec: clustered}). For a bin-to-bin comparison, the $P(D)$ prediction is interpolated onto the flux-density bins of the observed histogram defined in the last paragraph. We then evaluate the bin-to-bin difference using the likelihood function (Equation~\ref{eq:chi}).

For the MCMC sampling process, the initial state of the positions of the walkers uses the interpolated nodes on the real galaxy counts in the simulation catalogue with the addition of a Gaussian scatter to generate some randomness in the initial states for different walkers that satisfy the conditions mentioned in Section~\ref{ch:Models}. The problem becomes multimodal for different proposed models, and so we adopt a weighted move, with the differential evolution proposal described in \citet{Nelson2014} of 80\% probability and that with the `snooker updater' in \citet{ter2008differential} of 20\% probability.

\begin{figure*}[htbp!]
\centering
    \includegraphics[width=0.47\textwidth]{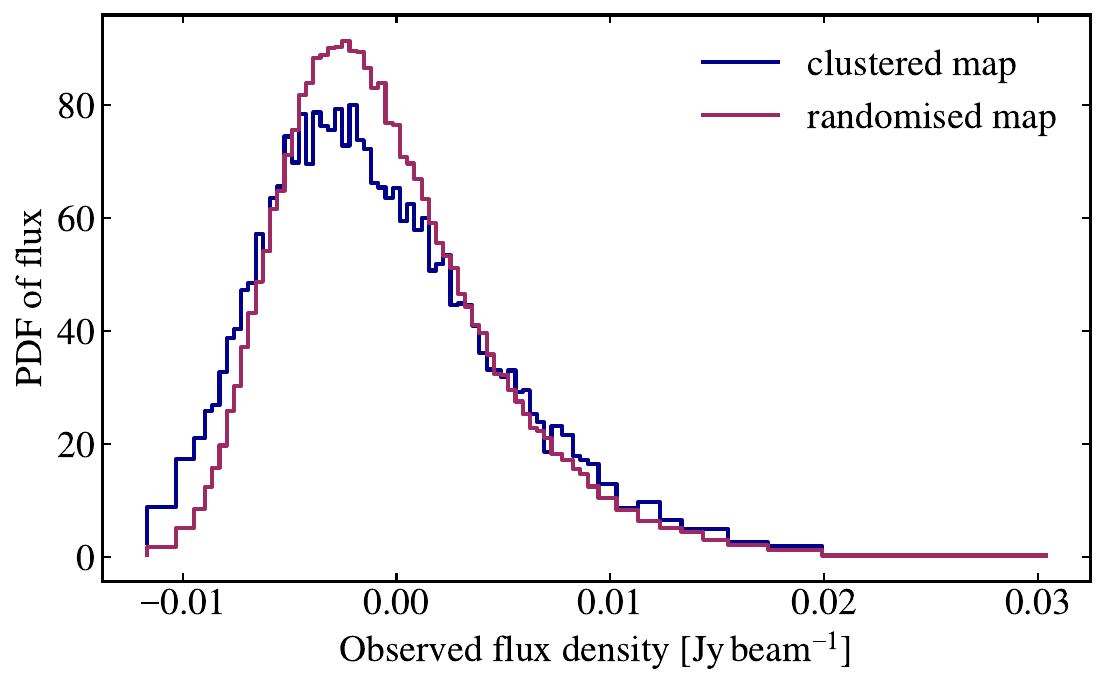}
    \includegraphics[width=0.48\textwidth]{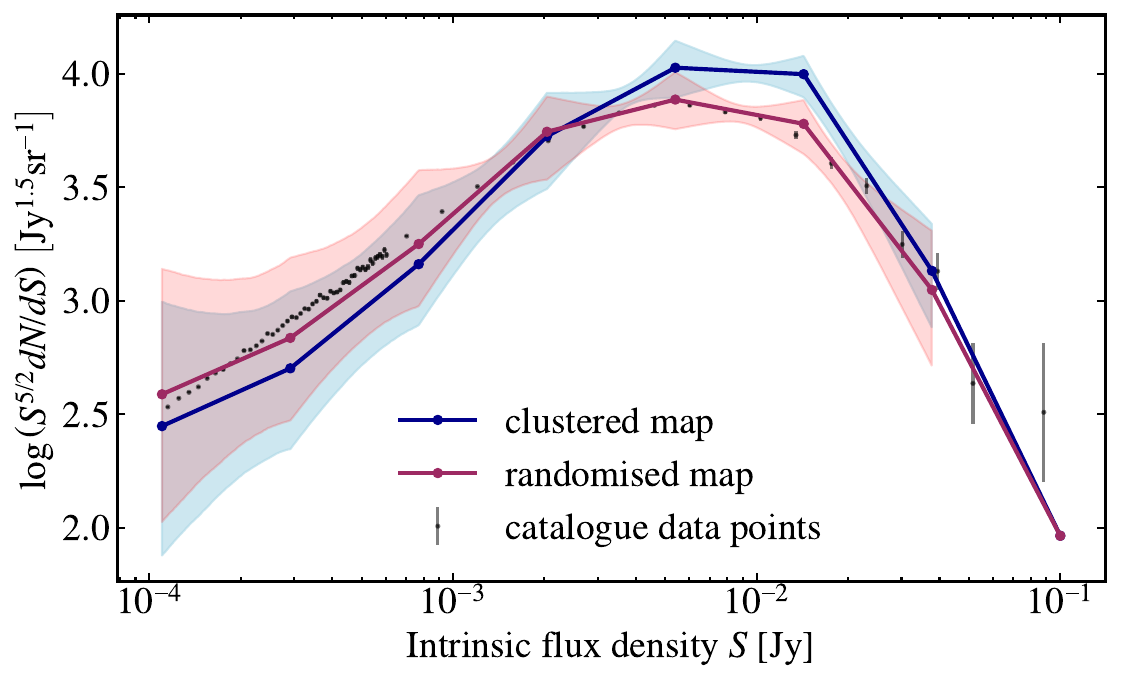}
    \caption{\textbf{Left panel: } Observed histograms of the clustered map (blue) and the randomised map (red), both generated from the simulated \textit{Herschel}-SPIRE map at 500$\,\mu$m. \textbf{Right panel:} $P(D)$ galaxy number counts fit to the clustered map (blue) and the randomised map (red). The galaxy number counts are Euclidean normalised here. The first node is an upper limit. The shaded region is the 68\% confidence region for each fit. The black dots are the real galaxy counts from the SIDES catalogue with Poisson errors in a finer binning for illustration.}
    \label{fig:pd}
\end{figure*}

The results of the fits on both the clustered map and the randomised map are shown in Figure~\ref{fig:pd}. The left panel shows the histograms, and the right panel shows the corresponding galaxy counts models that generate the best fit of the histograms in the left panel. The 68\% confidence region at each node is derived by obtaining the 16th and 84th percentiles of the samples after burn-in. The error bars in regions between the nodes are derived by interpolating 1000 uniform bins on the samples of the nine fit nodes to the whole flux range in each step. We thereby confirm that the galaxy counts fit is biased when directly applying $P(D)$ fluctuation analysis to the clustered map.

\section{Simulations: correction for clustering bias}
\label{ch:sim}

In Section~\ref{ch:Models}, we showed that the clustering of galaxies changes the flux histograms of observed maps, and thus biases galaxy number counts when using $P(D)$ fluctuation analysis. We now explore how clustering changes the histogram, i.e.\ the relation between clustering strength and the change in the flux histogram of clustered maps in Figure~\ref{fig:pd}.

We start by generating simulated maps with a range of different clustering strengths, but with the same underlying catalogue of galaxies from SIDES (Section~\ref{sec: clustered}). We then characterise the difference in the histograms caused by the clustering strength and provide a framework to correct the flux histogram, and thus the galaxy count fits (Section~\ref{sec: corr}), using the clustering strength, an observable that can be derived from the power spectrum of the map (Section~\ref{sec:ps}). The simulated mock observations in this section assumes the SPIRE 500-$\mu$m beam size, where the clustering bias should be the largest. 

\subsection{Generating maps of different clustering strength}
\label{sec: clustered}

Galaxy clustering can be characterised by the 2-point correlation function or the power spectrum of the map. Throughout this section, we discuss the effect of galaxy clustering only from 2-point statistics, since it is the main contributor to the change of the flux histogram compared to any higher-order correlations. In 2-dimensional source catalogues the 2-point correlation function $w(\theta)$ is defined as the excess probability of finding a second galaxy at an angular separation $\theta$ from a given galaxy. At small angular scales it is empirically approximated by a power law of the form 
\begin{equation}
w(\theta)=\left(\theta / \theta_{0}\right)^{-\alpha},
\label{eq:2pcf}
\end{equation}
where $\theta_0$ is the characteristic clustering scale and thus the parameterisation of the clustering strength. We assume that the power-law index is a constant, $\alpha = 0.8$, which has been found for a large number of optical galaxy surveys starting with \citet{DavisPeebles1983}, and is also found for radio galaxies \citep{BlakeWall2002a, BlakeWall2002b}. We expect modest deviations from this assumption not to significantly affect the correction form and we note that this is a simplification of actual galaxy clustering, and we will discuss the potential effect of this simplification in Section~\ref{sec:diss_clustering}.

\begin{figure*}[htbp!]
\centering
	\includegraphics[width=1.0\textwidth]{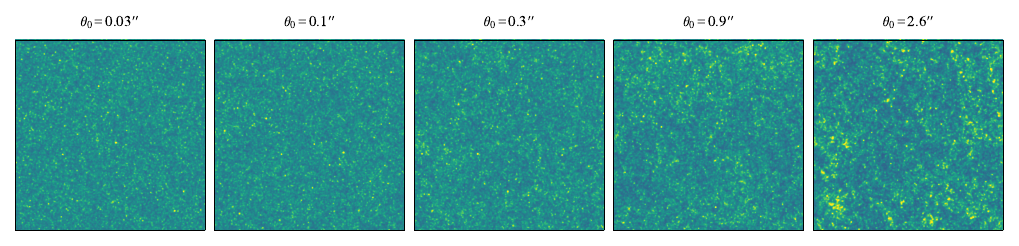}
    \caption{Examples of generated maps with different clustering strengths or characteristic clustering scales $\theta_0$ (Equation~\ref{eq:2pcf}), with exaggerated values for visualisation of the clustering effect. The map size is 2 $\text{deg}^2$, the same as the size of the SIDES simulation.}
    \label{fig:clustered}
\end{figure*}

\begin{figure*}[htbp!]
\centering
	\includegraphics[width=0.85\textwidth]{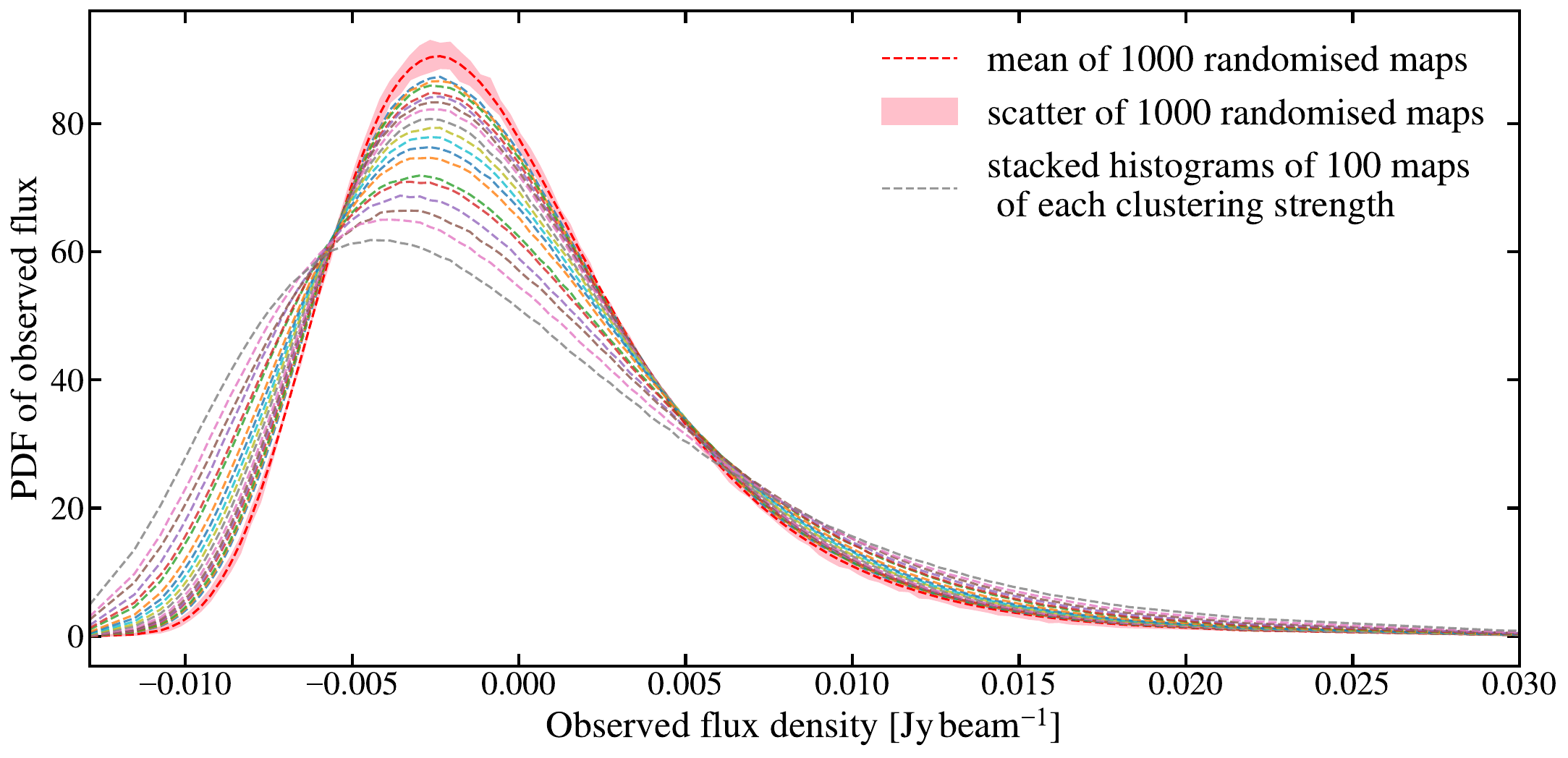}
    \caption{ Mean PDFs of 100 generated maps for each clustering strength or characteristic clustering scale $\theta_0$ (Equation~\ref{eq:2pcf}) in coloured dashed lines. The PDF is computed from the average histogram by Equation~\ref{eq:pdf}. The red dashed line and the shaded region represent the mean PDF and standard deviation of 1000 randomised maps. As the clustering increases, the peaks of the histograms shift to fainter flux densities and the tails grow.}
    \label{fig:histograms_shift} 
\end{figure*}

We generate modified maps from the SIDES galaxy catalogue with different characteristic clustering scales $\theta_0$, uniformly ranging from $0.036\, \mathrm{arcsec}$ to $1.38\, \mathrm{arcsec}$,\ using a method that we describe below. The range of $\theta_0$ is determined so that it is approximately centred at the actual clustering of the 2$\, \mathrm{deg}^2$ simulated map from SIDES in the same configuration. This clustering scale is found by comparing the histograms of the generated clustered maps and the actual SIDES map. For simplicity, we assume $\theta_0$ to be the same for all galaxies, regardless of their brightness.

We first generate, for different clustering strengths, the positions of all objects in the SIDES catalogue. There are several approaches that could be used to generate a density field that follows a designated 1D power spectrum. Here we assume a log-normal density field realisation, which is a good approximation for real cosmological density variations and a useful analytical form for simulation algorithms \citep[e.g.][]{ColesJones1991, Beutler2011}. 
For a given $\theta_0$, $w(\theta)$ can be converted to the angular power spectrum through the intrinsic Fourier relation \citep[see $\S$46 of][]{1980lssu.book.....P}. For a sky area of 2$\, \mathrm{deg}^2$ of interest here, we can adopt a flat-sky assumption and simplify the angular power spectrum to a 2D spatial Fourier transform of the map, following the convention used for CIB power spectra \citep[e.g.][]{Viero2013}. For $w(\theta)=\left(\theta / \theta_0\right)^{-\alpha} = A\theta^{-\alpha}$, where $A$ is the amplitude. In this relation we define $\theta$ to be in units of deg and $A$ in deg$^{\alpha}$.  The resulting clustering power spectrum is the Fourier transform of $w(\theta)$, 
\begin{equation}
 {\cal P}(k_\theta) =  \frac{2\pi A }{k_\theta^{2-\alpha}}\frac{2^{1-
\alpha} \Gamma((2-\alpha) / 2)}{\Gamma(\alpha / 2)},
\label{eq:2dft}
\end{equation}
where $k_\theta$ labels the Fourier modes in units of radians$^{-1}$. $ {\cal P}_\delta(k_\theta)$ is then in units of sr$^{-1}$. We sample $k$-space with 1000 modes and transform $ {\cal P}(k_\theta)$ to $ {\cal P}^\prime(k_\theta)$ using the relation of its covariance $w^\prime(\theta)=\ln(1+w(\theta))$, so that this lognormal realisation of $ {\cal P}^{\prime}(k_{\theta})$ is the same as the Gaussian realisation of $ {\cal P}(k_\theta)$ \citep[see][]{ColesJones1991}. We generate the lognormal density field for a $1200 \times 1200$ grid on a $2\, \mathrm{deg}^2$ map through the inverse Fourier transform of $ {\cal P}^\prime(k_\theta)$. We then draw Poisson samples from the lognormal density field for each cell plus a small random variation of the cell separation.

Next, we match flux densities of galaxies to the newly-generated positions by shuffling the flux densities to assure a uniform clustering of galaxies without any luminosity dependence. We assign galaxies with the new positions to a map with a pixel width of $12~\mathrm{arcsec}$ (as in the actual \textit{Herschel}-SPIRE maps at 500$\,\mu$m). We convolve the maps with the beam and add Gaussian noise with $\sigma=1.03\  \mathrm{mJy}~\mathrm{beam^{-1}}$, chosen for the same reason as mentioned in Section~\ref{sec:simulation-maps}. In Figure~\ref{fig:clustered}, we show examples of the generated clustered maps for five different clustering strengths to illustrate the effectiveness of the clustering algorithm. We adopt the adaptive binning set-up as in Section~\ref{sec:pdfit}. We generate 100 clustered maps of each clustering strength by shuffling the flux densities and matching them with the generated positions. We find the histograms of 100 clustered maps of each clustering strength and stack them together to find the average histogram for each clustering strength. For comparison, we run 1000 simulations, randomly assigning positions to galaxies in the SIDES catalogue, to produce maps in the same way, and obtain their stacked histograms and levels of scatter in each bin.

\begin{figure}[htbp!]
\centering
	\includegraphics[width=\textwidth]{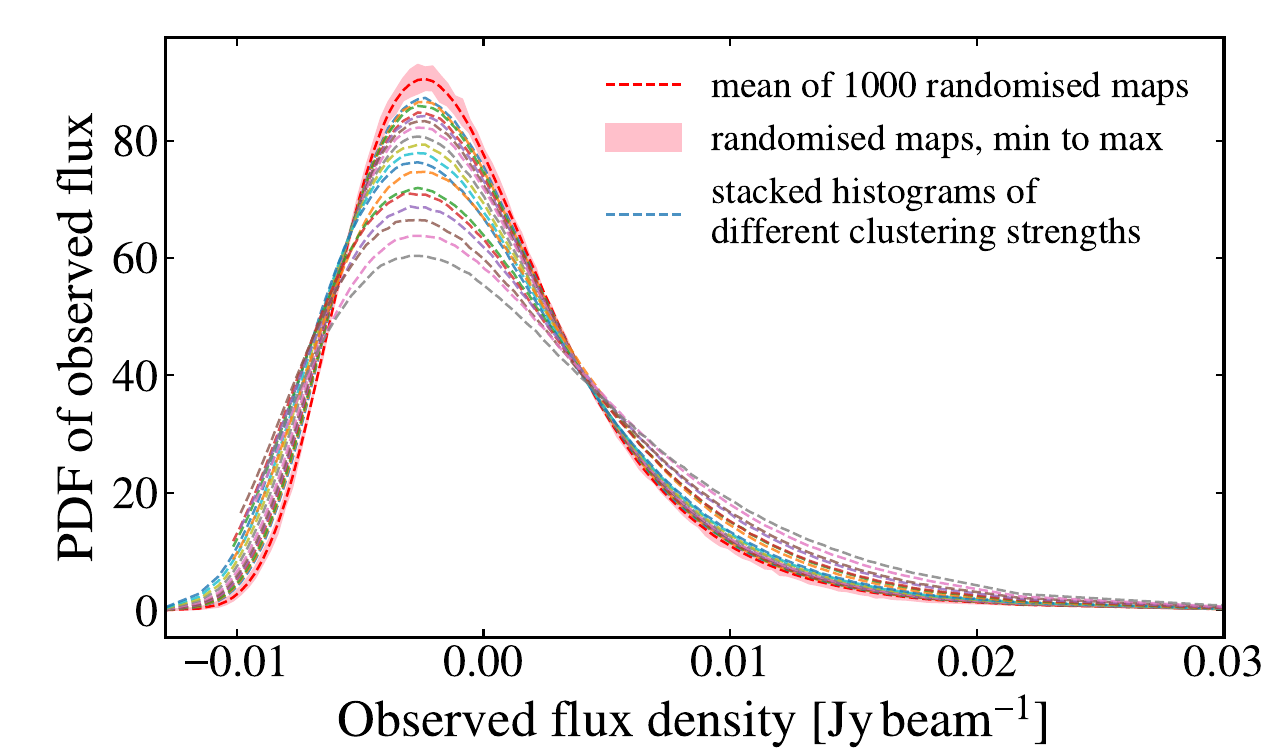}
 \includegraphics[width=\textwidth]{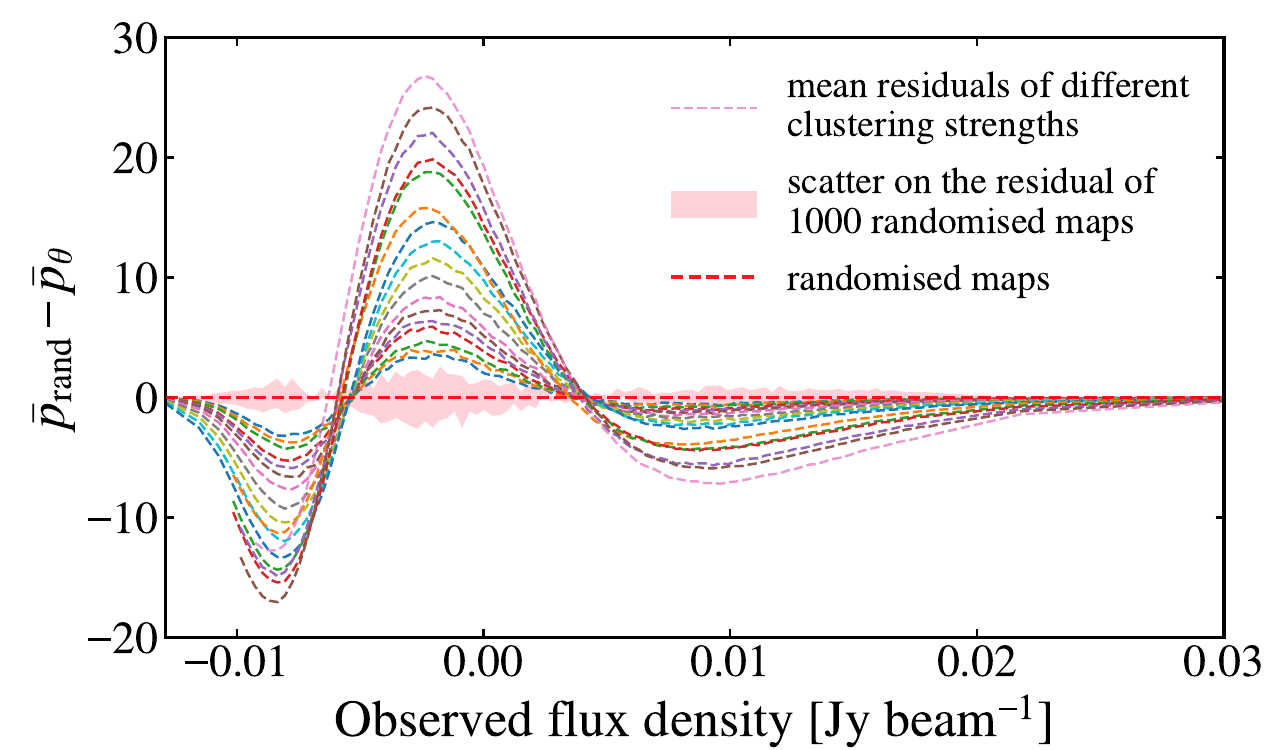}
    \caption{\textbf{Top Panel: }Same colour style as in Figure\,\ref{fig:histograms_shift}, but with the peaks of the histograms now aligned. \textbf{Bottom Panel: }The difference between the mean PDFs of the clustered maps and the randomised map, $\bar{p}_{\text{rand}} - \bar{p}_{\theta}$, for each clustering strength. The red dashed line and the shaded region are the zero-residual and scatter from 1000 randomised maps, respectively.}
    \label{fig:histograms} 
\end{figure}

The map histograms are then transformed into consistent PDFs following

\begin{equation}
p_i =  \frac{n_i}{N_{\textrm{pix}}\Delta x_i},
\label{eq:pdf}
\end{equation}

\noindent where $p_i$ is the PDF in the $i$th bin, $n_i$ is the number of pixels in the $i$th bin, $N_{\textrm{pix}}$ is the total number of pixels in the map, and $\Delta x_i$ is the bin width.

We show the resulting mean PDFs of different clustering strengths in the left of Figure\,\ref{fig:histograms_shift}. The peaks of the histograms move to fainter flux densities as the clustering strength grows larger.\footnote{We note that the shifts in the peaks of the histograms are linearly correlated with increasing clustering strength. However, we decide to fit for this shift in the counts-fitting stage instead of finding a deterministic relation with the clustering strength, due to the mean of the map being very sensitive to small shifts in the histogram.} We perform a simple spline fit in the peak regions and align the peaks of these PDFs. In Figure~\ref{fig:histograms}, we notice a clear trend that as the clustering becomes stronger, that is, the characteristic clustering scale $\theta_0$ increases, the peak of the histogram is suppressed and the tails are boosted. In other words, as galaxies get more clustered, there is an increasing number of brighter pixels (sources merging into pixels) and fainter pixels (because underdense regions are also created by clustering).

The effect seen in Figure~\ref{fig:histograms_shift} looks like a smoothly varying set of changes.  We will find it convenient to first shift the means (as described in the next section), which is shown in the upper panel of Figure~\ref{fig:histograms}.
 In the bottom panel of Figure~\ref{fig:histograms}, we show the difference, $\bar{p}_{\textit{rand}} - \bar{p}_{\theta}$, between the mean PDFs of randomised maps, $\bar{p}_{\textit{rand}}$ and those of clustered maps, $\bar{p}_{\theta}$ for each given $\theta_0$. 

We notice that the PDFs change in a systematic way, giving higher amplitude in the tails and a lower peak, representing an approximately single-parameter family of functions (and with the negative tail changing less significantly than the positive tail). Our goal is to create an asymmetric `W-shaped' correction function, whose amplitude we expect to depend on the clustering strength, or $\theta_0$. Then measuring the amplitude of the clustering should enable us to estimate how to correct the clustered histogram to provide the unclustered one, from which we can determine unbiased counts.

\subsection{Correction function fit on the stacked histogram}
\label{sec: corr}

We parameterise the correction function $F(x)$ by first aligning the peaks of the histograms found by simple spline fits and fitting $\bar{p}_{\text{rand}} - \bar{p}_{\theta}$.  We do this using a function combining a second-order polynomial and skewed Gaussian on each side of the peak,

\begin{equation}
F(x)=
\begin{cases}
A_1\sigma_2 Q(x,\mu_1,\sigma_1)\, \\
\qquad \times \operatorname{skew}\!\left(x,\sigma_2,\mu_2,a_1\right),
& x<x_{\rm peak},\\[3pt]
A_2\sigma_4 Q(x,\mu_3,\sigma_3)\, \\
\qquad \times \operatorname{skew}\!\left(x,\sigma_4,\mu_4,a_2\right)+k,
& x>x_{\rm peak}.
\end{cases}
\label{eq:corr0}
\end{equation}

\noindent Here $Q(x, \mu,\sigma) = 1-\left(\frac{x-\mu}{\sigma_{}}\right)^{2}$ and $\operatorname{skew}\left(x, \sigma, \mu, a\right)$ represents a skewed normal distribution with expectation value $\mu$, variance $\sigma^2$, and skewness $a$. 

\begin{figure}[htbp!]
\centering
	\includegraphics[width=\linewidth]{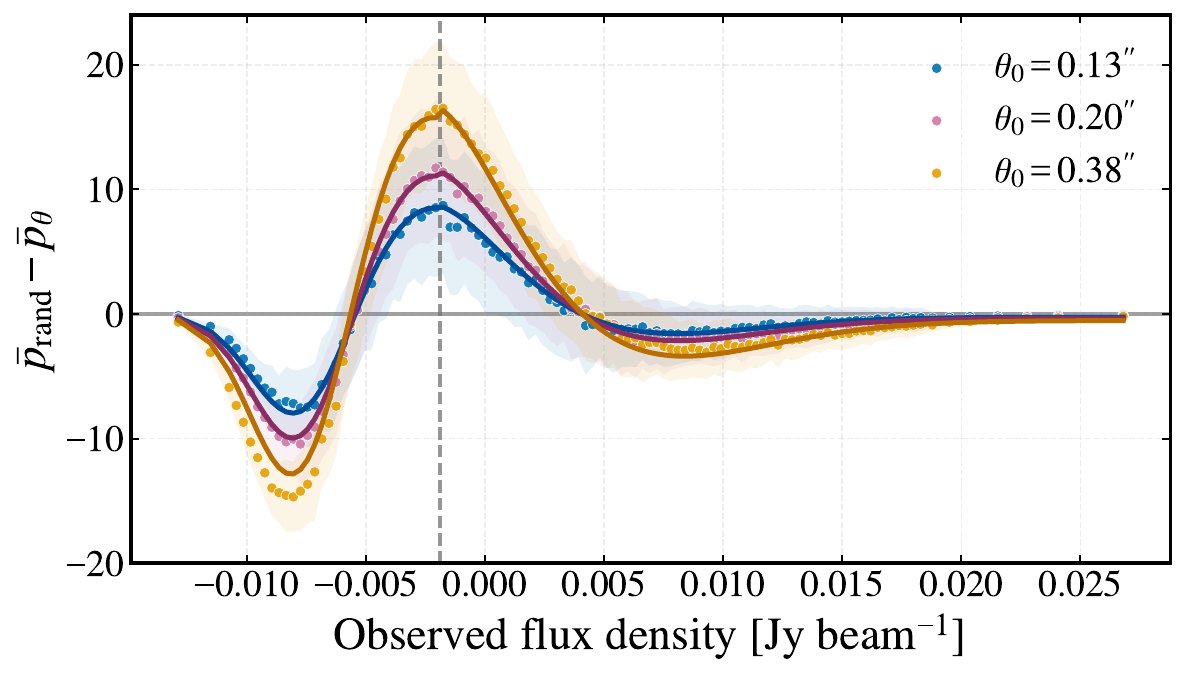}
    \caption{Examples of $\bar{p}_{\textit{rand}} - \bar{p}_{\theta}$ at different clustering strengths or $\theta_0$ (dots), with their respective fits from Equation~\ref{eq:corr0} in solid lines. The vertical dashed line represents the peak $x_\text{peak}$ in the equation. The shaded region represents the scatter in $\bar{p}_{\text{rand}} - \bar{p}_{\theta} $ in 100 simulated map pairs.}
    \label{fig:res_fit}
\end{figure}

\begin{table}[htbp!]
\centering
\resizebox{\textwidth}{!}{%
\begin{tabular}{@{}ccl@{}}
\toprule
\noalign{\hskip -3pt}
Parameters & Fixed value & \multicolumn{1}{c}{Fit relation}                               \\ \midrule
$A_1$ &  & \begin{tabular}[c]{@{}l@{}}$\displaystyle \begin{aligned}
\log A_1 &= (-0.089 \pm 0.02)\log^2 \theta_0 \\
         &\quad + (-0.02 \pm 0.20) \log \theta_0 \\
         &\quad -(5.21 \pm 0.44)
\end{aligned}$\end{tabular} \\
$\mu_1$    & $3.48$        &                                                                \\
$\sigma_1$ & $0.80$        &                                                                \\
$A_2$      &  & $\displaystyle \log A_2 = (0.78 \pm 0.02) \log \theta_0 + (-3.38 \pm 0.07)$  \\
$\mu_2$    &  & $\displaystyle \log \mu_2 = (0.78 \pm 0.03) \log \theta_0 + (0.32 \pm 0.10)$ \\
$\sigma_2$ & $0.74$        &                                                                \\
$\sigma_3$ &  & $\displaystyle \sigma_3 = (553.26 \pm 22.59) \theta_0 + (1.12 \pm 0.01)$     \\
$\mu_3$    &  & $\displaystyle \mu_3 = (3247.44 \pm 156.37) \theta_0 + (1.96 \pm 0.04)$      \\
$a$        & $1.03$        &                                                                \\
$k$        &  & $\displaystyle b = (-3560.90 \pm 87.87) \theta_0 + (-0.01 \pm 0.01)$         \\ \bottomrule
\end{tabular}%
}
\caption{Parameters in the correction function (Equation~\ref{eq:corr0}), either fixed or fit from a simple relation to the characteristic clustering scale $\theta_0$ in units of degree (Equation~\ref{eq:2pcf}). We note that since we simulated SPIRE 500-$\mu$m maps with noise levels matching existing GOODS-N observations, the parameters in this table are applicable to this particular data set (see Section~\ref{ch:goodsn}).}
\label{tab:p}
\end{table}

After aligning the histogram peaks to the average histogram of randomised maps, we then adopt the Levenberg-Marquardt least-squares method \citep[$\S$15.5 in][]{NumRec2007} to fit $\bar{p}_{\text{rand}} - \bar{p}_{\theta}$ with Equation~\ref{eq:corr0}, using equally-weighted bins to better recover the shape of the correction function. We explore the set of free parameters introduced in Equation~\ref{eq:corr0}, and find that some of them can be fixed to their mean values, i.e.~independent of $\theta_0$, while the other parameters can be fit by simple linear or log-linear relations with the characteristic clustering scale $\theta_0$. The best-fit relations for all parameters are shown in Table~\ref{tab:p}. We note that the amplitudes of $\bar{p}_{\text{rand}} - \bar{p}_{\theta}$ vary with the beam size and noise level of the map, so the parameters in the correction function are treated case by case at this stage. In this section we have adopted the SPIRE 500-$\mu$m beam and the GOODS-N noise level in this band, so the set of parameters in Table~\ref{tab:p} is readily applicable for this particular field and band and is used later in Section~\ref{ch:goodsn}. 

In Figure~\ref{fig:res_fit} we show the correction function overlaid with $\bar{p}_{\text{rand}} - \bar{p}_{\theta}$ after fixing and fitting the simplified relation to $\theta_0$.  The correction function follows the shape of the residuals well and we show later in Section~\ref{sec:results} that it sufficiently characterises the key features in the change of the histogram shapes, so that it almost completely debiases the $P(D)$ galaxy counts from the effects of galaxy clustering.

\subsection{Two-point correlation function and power spectra}
\label{sec:ps}
To determine the level of correction needed in the clustered map, we need to obtain the characteristic clustering scale, $\theta_0$.
The pointwise 2-point correlation function $w(\theta)$ is useful when there is a catalogue of galaxies in the observed field; however, the sources in a catalogue are usually flux-limited. For our application, we care about the clustering of relatively faint sources, since these are what bias the $P(D)$ prediction as in Section~\ref{sec:pdfit}. This means that the map-based power spectrum is a better place to focus our attention.

As mentioned in Section~\ref{sec: clustered}, the clustering power spectrum, $ {\cal P} (k)$, is the Fourier transform of $w(\theta)$, which describes the angular clustering of the source distribution. The angular power spectrum of an observed or simulated flux-intensity map is instead the 2D Fourier transform of the map intensity fluctuations in azimuthally-averaged Fourier component bins $k$:
\begin{equation}
\centering
P(k)=\left\langle|\delta I(\vec{k})|^{2}\right\rangle,
\label{ps_al}
\end{equation}
where $\delta I(\vec{k})$ is the intensity fluctuation field in Fourier space. 
Under the assumption that source fluxes are uncorrelated with source positions, the map power spectrum can be written as
\begin{equation}
\centering
P(k)=P_\mathrm{shot} + \bar I^2  {\cal P}(k),
\label{eq:ps_2}
\end{equation}
where $\bar I$ is the mean map intensity, $\bar I = \Sigma_i{S_i}/\Omega$ for a source catalogue with galaxy fluxes $S_i$ over a total angular area $\Omega$ and
$P_\mathrm{shot}$ is the contribution from Poisson shot noise.

We use the simulated clustered maps, without convolving with the beam or adding noise, so that we can calculate power spectra directly of the maps without any corrections for masking, beam functions, transfer functions, etc. We calculate the power spectra by taking the 2D Fourier transform at a sampling rate of the pixel width. We divide $k$-space into $2^8$ bins, sum over the power in each $k$ bin, and normalise the power spectrum by the number of pixels in each $k$ bin.

\begin{figure}[htbp!]
\centering
	\includegraphics[width=\textwidth]{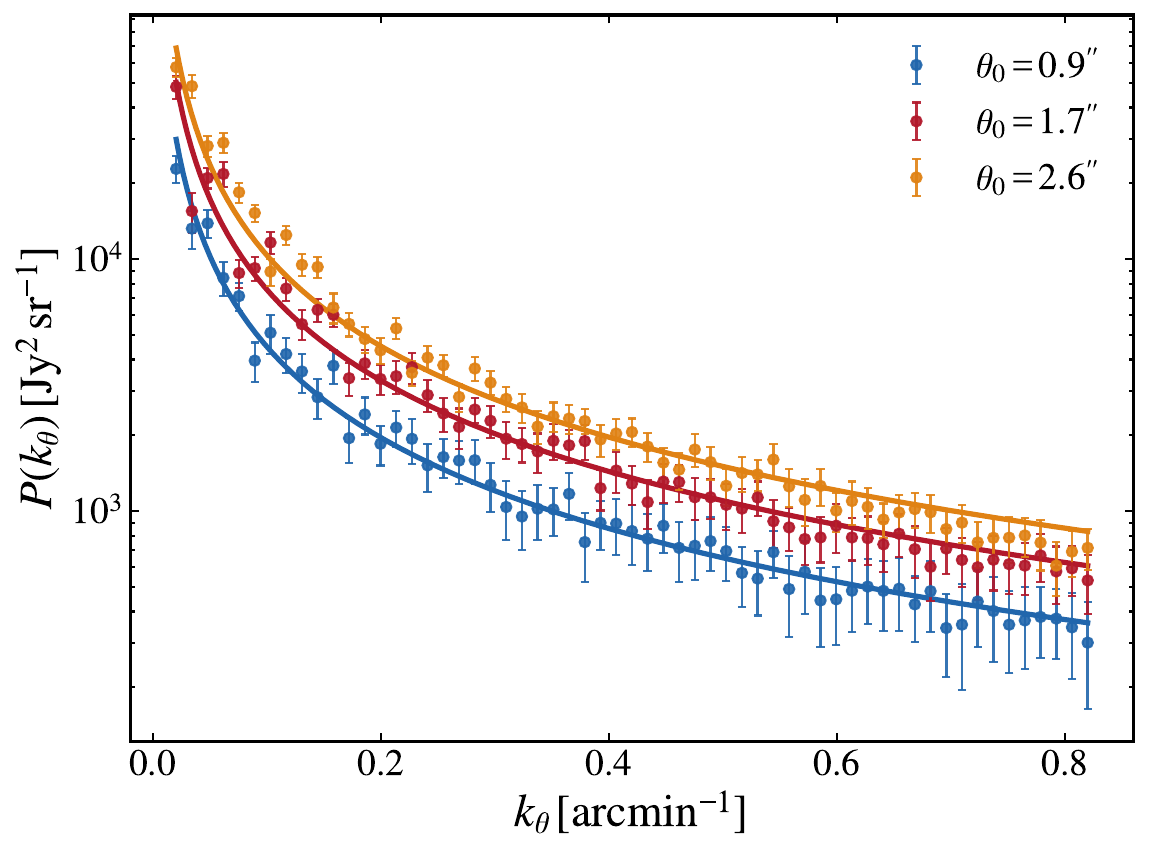}
    \caption{Examples of power spectra for generated maps at 500\,$\mu$m with different clustering strengths or characteristic clustering scales $\theta_0$ (Equation~\ref{eq:2pcf}). The constant shot noise term has been subtracted from these power spectra. The solid line is given by Equation~\ref{eq:log-linear}.}
    \label{fig:ps}
\end{figure}

We calculate the power spectra for 100 maps of each clustering strength and stack the power spectra in each $k$-bin. We then remove the constant shot-noise component ($P_\mathrm{shot}$\footnote{In principle this is related to the counts \cite[see e.g.][]{ScottWhite1999}, and so is an effect of the 1-point function on the 2-point function.  This could be explored in the future, but for now we treat the amplitude of the shot-noise power as a free parameter to be fit.}) from the power spectra by fitting a constant in the range $1.93\, \text{arcmin}^{-1} \leq k_\theta < 3.00\, \text{arcmin}^{-1}$, where $1.93 \,$arcmin$^{-1}$ corresponds to the scale of the beam FWHM, and $3 \,$arcmin$^{-1}$ is set by the map size. The resulting power spectra can be written as the power-law relation

\begin{equation}
\centering
\log \left(P(k_\theta) \right)=-1.2\log(k_\theta)+C,
\label{eq:log-linear}
\end{equation}
where $P(k_\theta)$ is the amplitude of the power spectrum in units of Jy$^2$\,sr$^{-1}$ and $k_\theta$ the Fourier modes in units of $\text{arcmin}^{-1}$. The slope $-1.2$ is fixed from the slope of the 2-point correlation function $w(\theta)$ by Equation~\ref{eq:2dft} and $C$ is determined by the relation

\begin{equation}
\begin{aligned}
C &= 0.8 \log(\theta_0)+ C_0, \\
\end{aligned}
\label{eq:A}
\end{equation}
where $\theta_0$ is in units of deg. The slope 0.8 is similarly set by the power-law index of $w(\theta)$. The intercept $C_0$ is set by the catalogue and can be analytically calculated by substituting $A\,{=}\,\theta_0^{\alpha}$ into Equation~\ref{eq:2dft} and noting that $P_\mathrm{shot}$ in $P(k)$ has already been subtracted from Equation~\ref{eq:log-linear}. We find $C_0\,{=}\,$6.02, 5.82 and 5.39 at 250, 350, and 500\,$\mu$m, respectively. We show the example power spectra and the fits at 500\,$\mu$m in Figure~\ref{fig:ps}.

By inverting Equation~\ref{eq:A}, we can estimate the characteristic clustering scale $\theta_0$ from the map power spectrum and find the appropriate form of the correction function (Equation~\ref{eq:corr0}) from the parameters in Table~\ref{tab:p}. We note that in this section we focus on testing the correction at 500\,$\mu$m, while the different clustering relations for 250 and 350\,$\mu$m shown in Equation~\ref{eq:A} are quoted later in Section~\ref{ch:goodsn}. We now have the 2-point information on spatial clustering, and we can propagate its impact on the flux histograms (i.e.\ 1-point statistics) via the correction function, and then recover the unbiased galaxy number counts using the $P(D)$ fluctuation analysis. In this way, we simultaneously obtain the 1-point (flux histograms or PDFs) and 2-point statistics (power spectrum or 2PCF) contained in a map, which together comprise nearly all the information available in an unresolved galaxy imaging survey.

\subsection{Results for corrected galaxy number counts}
\label{sec:results}

\begin{figure*}[htbp!]
\centering
	\includegraphics[width=\textwidth]{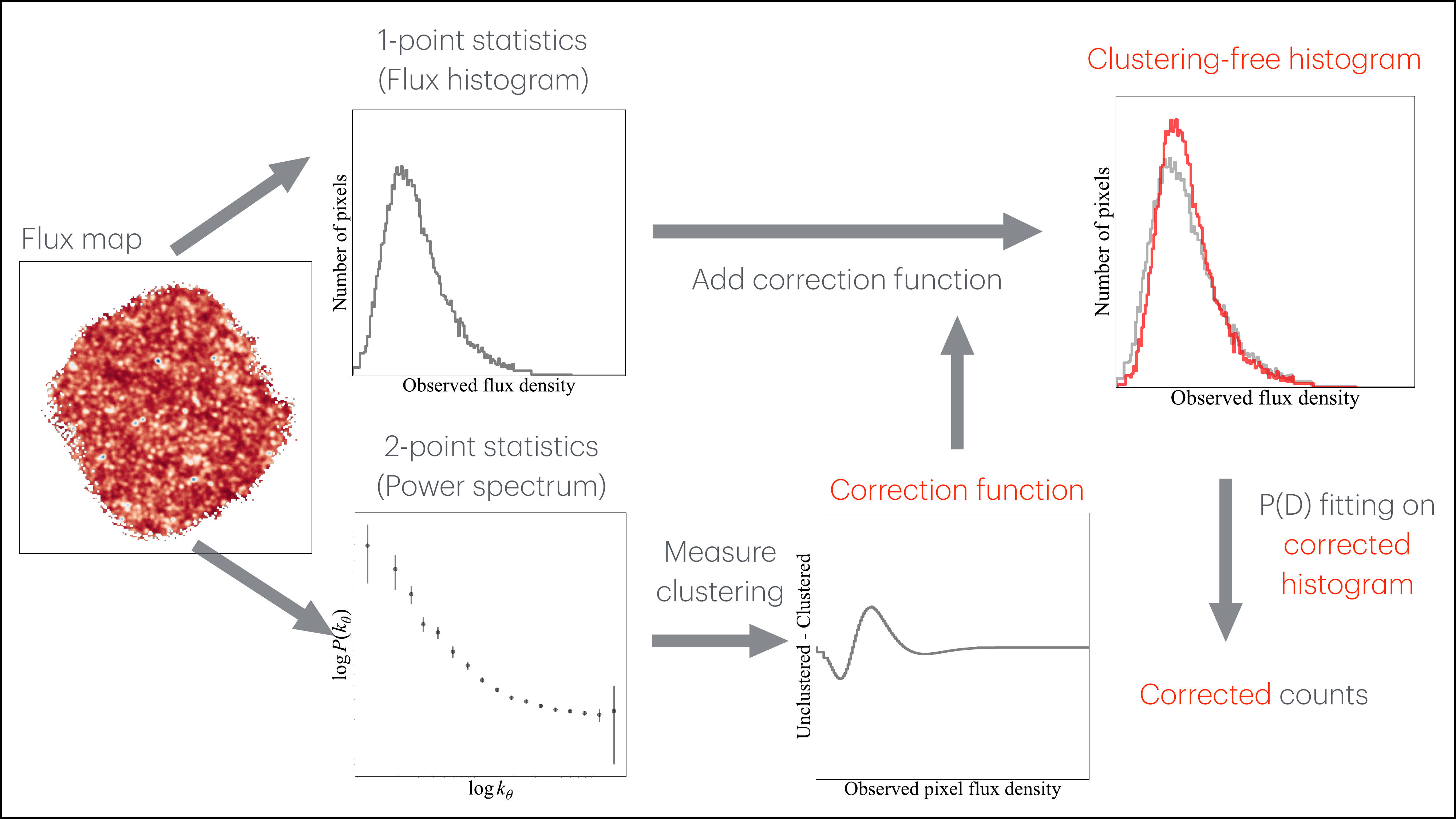}
    \caption{Flow chart illustrating the process for correcting the clustering bias in galaxy number counts fitting through $P(D)$ fluctuation analysis. Since $P(D)$ fits the flux histogram of the map, the process focuses on removing the impact of clustering (2-point statistics) on the flux histogram (1-point statistics). The specific form used in the correction is derived from simulations and the amplitude of clustering from the data themselves.}
    \label{fig:flow}
\end{figure*}

\begin{figure*}[htbp!]
\centering
	\includegraphics[width=0.49\textwidth]{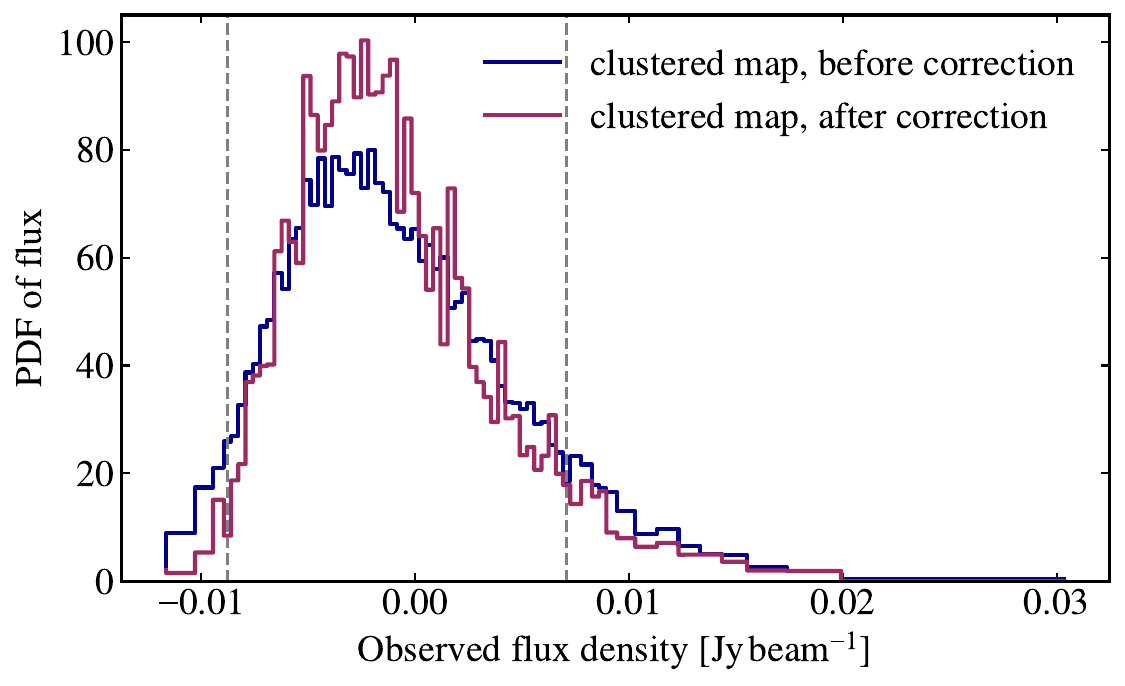}
    \includegraphics[width=0.49\textwidth]{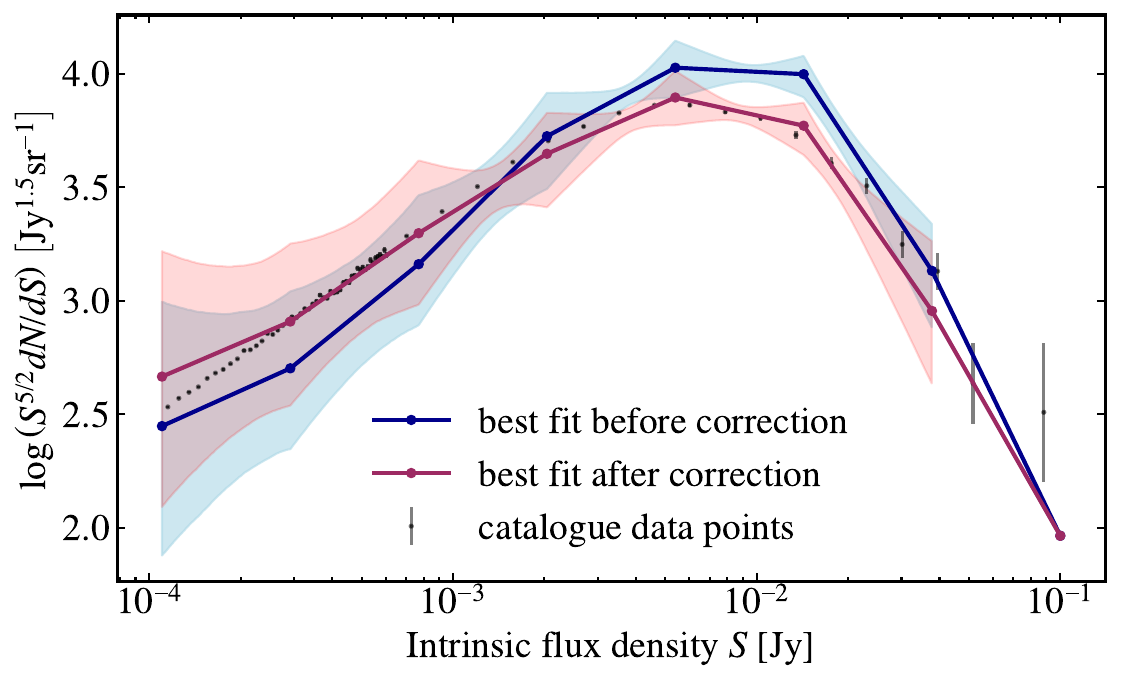}
    \caption{\textbf{Left panel: }Observed histograms of the clustered map before (blue) and after (red) correction in solid lines, after aligning their peaks. The grey dashed lines are the flux cutoffs beyond which the data are not used in the fit. \textbf{Right panel:} $P(D)$ galaxy number counts models before (blue) and after (red) correction. The galaxy number counts are Euclidean normalised here. The first node is an upper limit. The shaded band is the 68\% confidence region for each fit. The black dots are the real galaxy counts from the SIDES catalogue with Poisson errors in an arbitrary binning for illustration.}
    \label{fig:corr}
\end{figure*}

We summarise the correction process in a flow chart (Figure~\ref{fig:flow}). 
The usual $P(D)$ framework is built exclusively on fitting the flux histogram, while the impact of clustering is not considered, leading to biases in the estimated galaxy number counts. However, we can determine the clustering strength from the data themselves and then correct the flux density histogram to find what it would be for unclustered sources. $P(D)$ analysis conducted on this corrected histogram should therefore result in unbiased galaxy number counts.

To test our correction method, we use a simulated clustered map with a similar clustering strength as in the SIDES simulation, $\theta_0 = 0.33^{\prime\prime} \pm 0.003^{\prime\prime}$ calculated from Equation~\ref{eq:log-linear} after fitting the SIDES simulated map power spectrum by Equation~\ref{eq:log-linear}. We follow the correction processes as summarised in Figure\,\ref{fig:flow}.
We estimate the power spectrum using Equation~\ref{ps_al} and fit it with Equation~\ref{eq:log-linear} after subtracting $P_\text{shot}$, which is fit as a constant in the 
flat (high $k$) part of $P(k)$.
The uncertainty in the power spectrum is the dispersion of the power in each $k$ bin. We then find the characteristic clustering scale $\theta_0$ by inverting Equation~\ref{eq:A}.
We insert $\theta_{0, \text{fit}}$ into the relations in Table~\ref{tab:p}, where we find all the parameters needed to calculate the correction function in Equation~\ref{eq:corr0}. We add the correction function to the PDF of the map histogram, with the binning created adaptively as in Section~\ref{sec: clustered}. The corrected histogram is obtained by converting from the corrected PDF following Equation~\ref{eq:pdf}.

For the uncertainty in the characteristic clustering scale, $\theta_0$, we draw samples from a normal distribution within its $1\sigma$ error, estimated from Equation~\ref{eq:A} after fitting the power spectrum by Equation~\ref{eq:log-linear}. We then find the error bar of the correction function by drawing 20{,}000 samples of each variable coefficient in Table~\ref{tab:p} for each $\theta_0$ draw within its covariance range, and taking the 16th and 84th percentiles of the resulting correction function. To ensure numerical stability in the positive tail, we apply a mild winsorisation to the ensemble of correction function realisations, clipping extreme outliers at the 1st and 99th percentiles before estimating the uncertainty envelope. We assume the error bar to be symmetric, so $\sigma_\text{corr, sample} = (F_{\text{corr, sample}, 84} - F_{\text{corr, sample}, 16})/2$. We then add in quadrature a 10\% uncertainty ($\sigma_\text{sys} = 10\%~\sigma_\text{corr}$) to account for systemetics such as small beam variations (see Section~\ref{sec:sys}). The total uncertainty in the correction function is then $\sigma_\text{corr} = (\sigma_\text{corr, sample}^2 + \sigma_\text{sys}^2)^{1/2}$. This is included in the likelihood function (Equation~\ref{eq:chi}), where $\sigma_i$ becomes 

\begin{equation}
\centering
\sigma_i =\sqrt{\sigma_\text{Poisson}^2+\sigma_\text{corr}^2} = \sqrt{n_i + \sigma_\text{corr}^2},
\end{equation}

\noindent while for the standard process without using a correction function, included later for comparison, we use $\sigma_i = \sigma_\text{Poisson}$.

Following this procedure, we show in Figure~\ref{fig:corr} the flux histograms before and after correction, and the corresponding final galaxy counts fit in the same fashion as in Section~\ref{sec:pdfit}. The corrected counts clearly recover the true underlying counts much better than the fit to the uncorrected counts.

To evaluate the goodness of fit of the galaxy counts in comparison to the real catalogue, we first bin the galaxies in the catalogue according to their brightness, with logarithmic bins at the brighter end to ensure there are more than 10 galaxies in each bin. The bins are extra fine here
to fully trace the shape of the underlying counts, especially at the faint end. We then transform the number of galaxies in each bin to the galaxy number counts by dividing it by the bin size and the total area of the map. The uncertainty in the resulting galaxy counts from the catalogue is evaluated by taking the errors in each bin to be $\sqrt{n_i}$, assuming a Poisson distribution of the number of galaxies $n_i$ in the $i$th bin, and propagating the error into the number counts, $\sigma_\text{cat}$. The binned counts then represent the true galaxy counts in SIDES.

We can evaluate the goodness of fit of the galaxy counts by calculating $\chi^2$. The fit nodes of the galaxy counts are interpolated by power laws onto the central bin values. We then calculate

\begin{equation}
\centering
 \chi^{2}=\sum_{i} \frac{\left(D_{i}-M_{i}\right)^{2}}{\sigma_{i}^{2}},
\label{eq:chi_r}
\end{equation}
where $D_i$ is the galaxy count in the $i$th bin calculated from the simulation catalogue, $\sigma_i$ is the error on $D_i$, $M_i$ is the interpolated counts in the $i$th bin from the fit results, and there are 805 degrees of freedom in the galaxy counts.
We find that $\chi^2$ drops from 10581 to 915 after the correction,
a dramatic improvement. This demonstrates that the uncorrected fit is systematically biased, and the proposed clustering correction dramatically improves the $P(D)$ galaxy number counts fit.
We have therefore shown that the proposed correction method is effective on simulated \textit{Herschel}-SPIRE 500-$\mu$m maps, with noise levels typical of GOODS-N deep observations.

\section{Application to Herschel-SPIRE data}
\label{ch:goodsn}

We now apply the correction method described above to the GOODS-N field, part of the HerMES and GOODS-\textit{Herschel} projects \citep{Elbaz2011,Oliver2012}. We use the full nested maps in the 4th release of HerMES, as part of the Herschel Extragalactic Legacy Programme (HELP; \citealt{Shirley2021}). This field was previously studied via $P(D)$ in \citet{Glenn2010} using the science-demonstration-phase (SDP) SPIRE data, while \citet{Bethermin2012} analysed the full GOODS-N and COSMOS data by stacking on IRAC 24-$\mu$m catalogues from \citet{Magnelli2011}. Here we present the first $P(D)$ analysis on the full-depth data, and with the impact of clustering corrected using the method described in Section~\ref{ch:sim}.

The parameters in the correction function are highly dependent on the beam and noise level of the observation, so a separate treatment is required for each SPIRE waveband. In the last section, we fit the parameters appropriate for the SPIRE data in the GOODS-N field at 500\,$\mu$m (Table~\ref{tab:p}). We then simulate observations with the beams and noise levels of the other two SPIRE bands at 250\,$\mu$m and at 350\,$\mu$m.
The clustering strength in GOODS-N is assumed to be the average of the power spectrum measurements determined by \citet{Viero2013} for \textit{Herschel}-SPIRE maps over $70 \,$deg$^2$ and we use this to estimate the correction function.

\subsection{GOODS-N field}

\subsubsection{Data}
\label{sec:G_data}

\begin{table}[]
\begin{tabular}{@{}ccccc@{}}
\toprule
SPIRE band                 & FWHM ('') & $\sigma_\text{inst}$ (mJy) & $\sigma_\text{inst, crop}$ (mJy) & $\sigma_\text{conf}$ (mJy) \\ \midrule
250\,$\mu$m & 18.15 & 1.85                          & 0.88                               & 5.8                         \\
350\,$\mu$m & 25.15 & 1.81                          & 0.87                               & 6.3                         \\
500\,$\mu$m & 36.3 & 2.15                          & 1.03                               & 6.8                         \\ \bottomrule
\end{tabular}
\caption{Specifications of the \textit{Herschel}-SPIRE GOODS-N maps \citep{Shirley2021}, including the beam size FWHM, mean instrumental noise $\sigma_\text{inst}$, mean instrumental noise of the cropped map used for this analysis $\sigma_\text{inst, crop}$, and the confusion noise $\sigma_\text{conf}$.} 
\label{tab:data}
\end{table}

The confusion noise for \textit{Herschel}-SPIRE is 5.8, 6.3, and 6.8$\,$mJy$\,$beam$^{-1}$ at 250, 350 and 500$\,\mu$m \citep{Nguyen2010} and the GOODS-N maps are well into the confusion-limited regime. Pixel sizes are 6, 8.33 and 12\,arcsec at 250, 350, and 500\,$\mu$m, respectively. The original observations span around $0.64\,$deg$^{2}$, but we cropped the map to retain the central area of approximately $0.3\,$deg$^2$ where the variation of the noise level is within 10\%.
The resulting noise levels of the cropped maps are shown in Table~\ref{tab:data}. The very deep SPIRE data in the GOODS-N field makes it ideal for studying the faint galaxy number counts. Although larger fields would effectively reduce field-to-field variance, their noise behaviour would be more complicated, requiring multi-zone $P(D)$ analyses \citep[e.g.][]{Vernstrom2014}, and the shallower depths of the other fields adds little constraining power at the faintest flux densities.

\subsubsection{Simulations and correction function}
\label{sec:cf_goodsn}

\begin{table}[htbp!]
\centering
\resizebox{\textwidth}{!}{%
\begin{tabular}{@{}ccl@{}}
\toprule
\noalign{\hskip -3pt}
Parameters & Fixed value & \multicolumn{1}{c}{Fit relation}                               \\ \midrule
$A_1$ &  & \begin{tabular}[c]{@{}l@{}}$\displaystyle\begin{aligned}
\log A_1 &= (-0.015 \pm 0.052)\log^2 \theta_0 \\
&\quad + (+0.49 \pm 0.47) \log \theta_0 + (-0.08 \pm 1.06)
\end{aligned}$\end{tabular} \\
$\mu_1$ &  & $\mu_1 = (-1.06 \pm 1.74) \theta_0 + (0.00 \pm 0.00) $ \\
$\sigma_1$ & $0.002$ & \\
$\mu_2$ & $-0.0063$ & \\
$\sigma_2$ & $0.39 $ & \\
$a_1$ & $409.89$ & \\
$A_2$ &  & \begin{tabular}[c]{@{}l@{}}$\displaystyle\begin{aligned}
\log A_2 &= (-0.15 \pm 0.04) \log^2 \theta_0 \\
&\quad + (-0.69 \pm 0.35) \log \theta_0 + (+158 \pm 0.74)
\end{aligned}$\end{tabular} \\
$\mu_3$ & $0.0027$ & \\
$\sigma_3$ &  & $\sigma_3 = (0.28 \pm 0.44) \theta_0 + (+0.034 \pm 0.00)$ \\
$\mu_4$ &  & $\mu_4 = (-1272.30 \pm 116.33) \theta_0 + (-1.96 \pm 0.020)$ \\
$\sigma_4$ &  & $\sigma_4 = (469.3 \pm 42.8) \theta_0 + (0.72 \pm 0.01)$ \\
$a_2$ & $1.41$ & \\
$k$ &  & $k = (-1740.18 \pm 87.84) \theta_0 + (0.000 \pm 1)$ \\
\bottomrule
\end{tabular}%
}
\caption{Same parameters as in Table~\ref{tab:p} for the GOODS-N map at 250\,$\mu$m (Equation~\ref{eq:corr0}).}
\label{tab:p_g_250}
\end{table}

\begin{table}[htbp!]
\centering
\resizebox{\textwidth}{!}{%
\begin{tabular}{@{}ccl@{}}
\toprule
\noalign{\hskip -3pt}
Parameters & Fixed value & \multicolumn{1}{c}{Fit relation}                               \\ \midrule
$A_1$ &  & \begin{tabular}[c]{@{}l@{}}$\displaystyle\begin{aligned}
\log A_1 &= (-0.20 \pm 0.021)\log^2 \theta_0 \\
&\quad + (-1.14 \pm 0.19) \log \theta_0 + (3.00 \pm 0.40)
\end{aligned}$\end{tabular} \\
$\mu_1$ &  & $\mu_1 = (-2.28 \pm 0.16) \theta_0 + (12.55 \pm 0.00)$ \\
$\sigma_1$ & $12.56$ & \\
$\mu_2$ & $-0.0083$ & \\
$\sigma_2$ & $0.004$ & \\
$a_1$ & $2.87$ & \\
$A_2$ &  & \begin{tabular}[c]{@{}l@{}}$\displaystyle\begin{aligned}
\log A_2 &= (-0.19 \pm 0.02) \log^2 \theta_0 \\
&\quad + (-1.00 \pm 0.14) \log \theta_0 + (+156.78 \pm 0.30)
\end{aligned}$\end{tabular} \\
$\mu_3$ & $0.0083$ & \\
$\sigma_3$ &  & $\sigma_3 = (1.12 \pm 0.01) \theta_0 + (+0.00 \pm 0.00)$ \\
$\mu_4$ &  & $\mu_4 = (-1807.41 \pm 0.36) \theta_0 + (-2.39 \pm 0.00)$ \\
$\sigma_4$ &  & $\sigma_4 = (67.07 \pm 0.01) \theta_0 + (0.08 \pm 0.00)$ \\
$a_2$ & $1.42$ & \\
$k$ &  & $k = (-1143.02 \pm 68.19) \theta_0 + (-0.04 \pm 0.00)$ \\
\bottomrule
\end{tabular}%
}
\caption{Same parameters as in Table~\ref{tab:p} for the GOODS-N map at 350\,$\mu$m (Equation~\ref{eq:corr0}).}
\label{tab:p_g_350}
\end{table}

Similarly to Section~\ref{ch:sim}, we simulate 1000 randomised maps and 100 clustered maps of clustering strengths from 2.06\, arcsec to 63.61\,arcsec, excluding unrealistically strong clustering. The generated maps are 2~deg$^2$, larger than GOODS-N to reduce the error after averaging the histogram.
We then fit $\bar{p}_{\text{rand}} - \bar{p}_{\theta}$ with the correction function form in Equation~\ref{eq:corr0} in a similar manner as in Section \ref{sec: corr}, and show the parameters of the correction function for 250\,$\mu$m and 350\,$\mu$m in Tables~\ref{tab:p_g_250} and \ref{tab:p_g_350}, respectively.

We take the auto-power spectrum at 250, 350 and 500\,$\mu$m from \citet{Viero2013}, choosing the deepest flux cut at $50~\mathrm{mJy}$ for the power spectrum. We then fit Equation~\ref{eq:log-linear}, finding the intercepts $C= 2.64$, $2.54$, and $2.18$, and the corresponding $\theta_\mathrm{0} = 0.26^{\prime\prime} \pm 0.002^{\prime\prime}$, $0.37^{\prime\prime} \pm 0.004^{\prime\prime}$, and $0.43^{\prime\prime} \pm 0.004^{\prime\prime}$ for 250, 350, and 500\,$\mu$m, respectively, through Equation~\ref{eq:A}.
We note that these $\theta_0$ values are derived from an empirical relationship, assuming $w(\theta)$ follows a single power law.

\begin{figure}[htbp!]
\centering
\includegraphics[width=\textwidth]{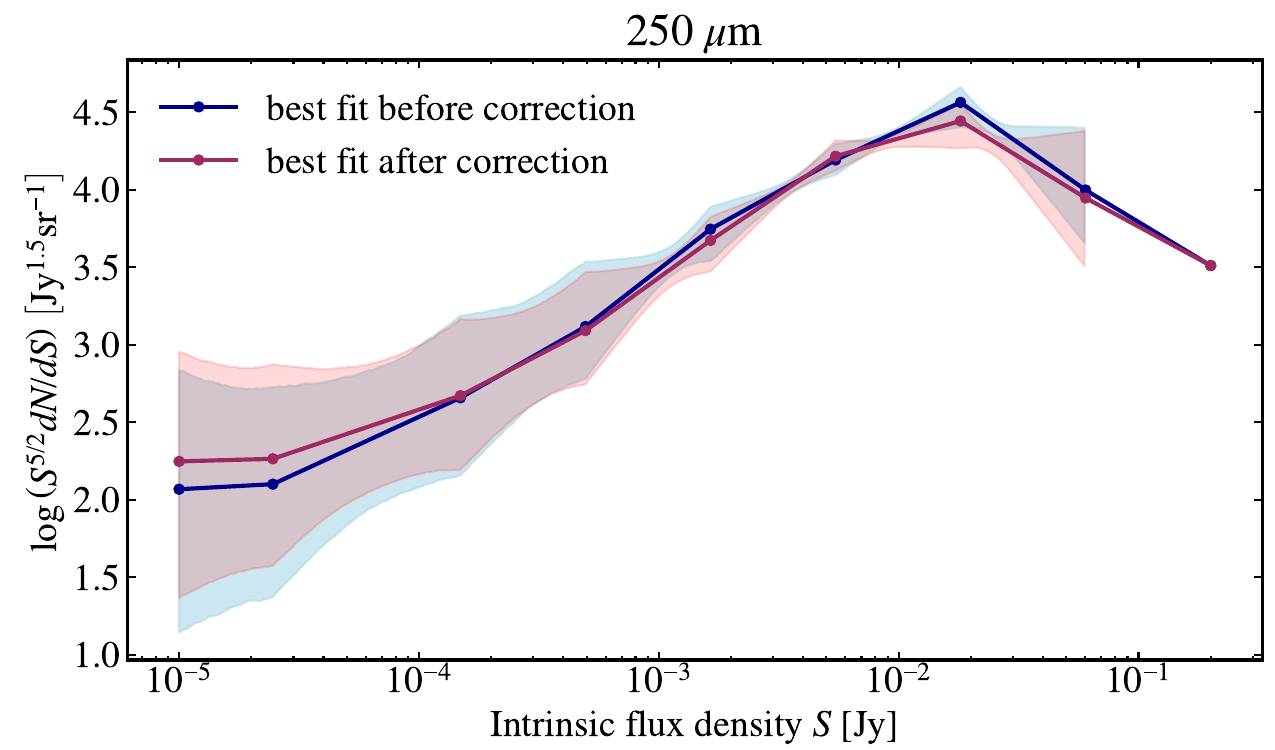}
\includegraphics[width=\textwidth]{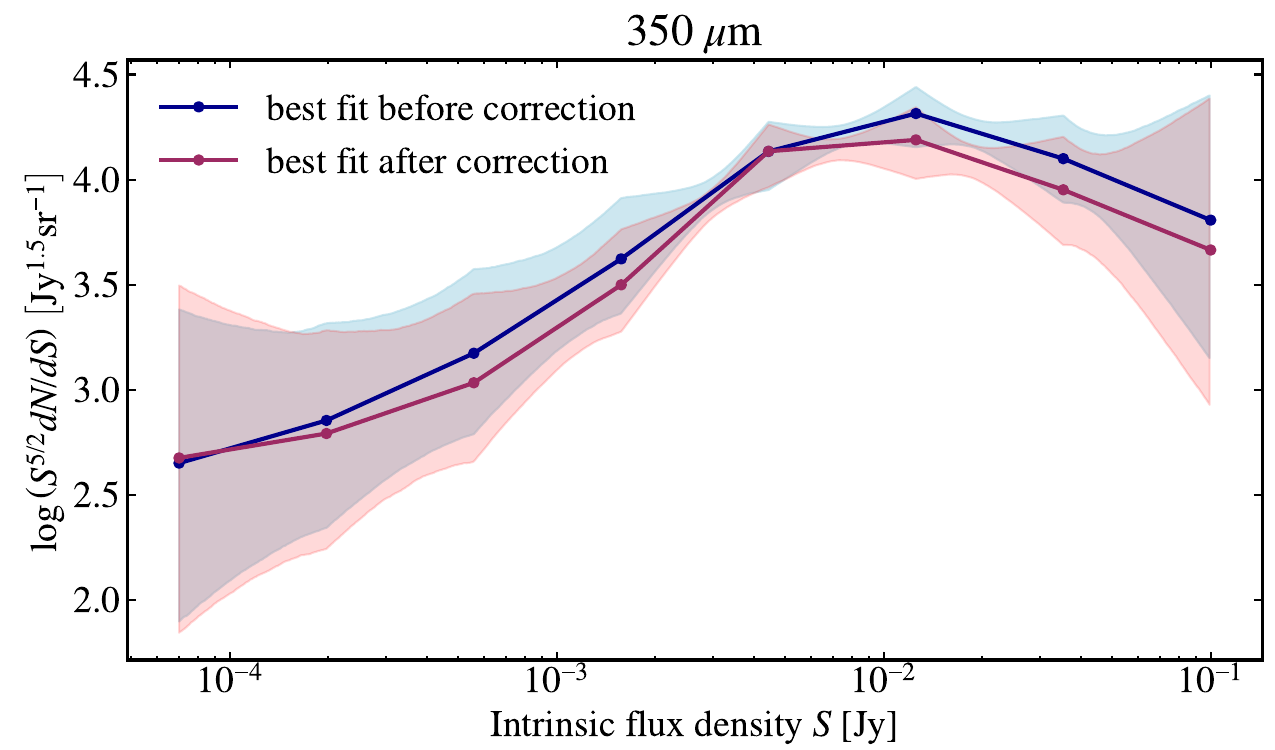}
\includegraphics[width=\textwidth]{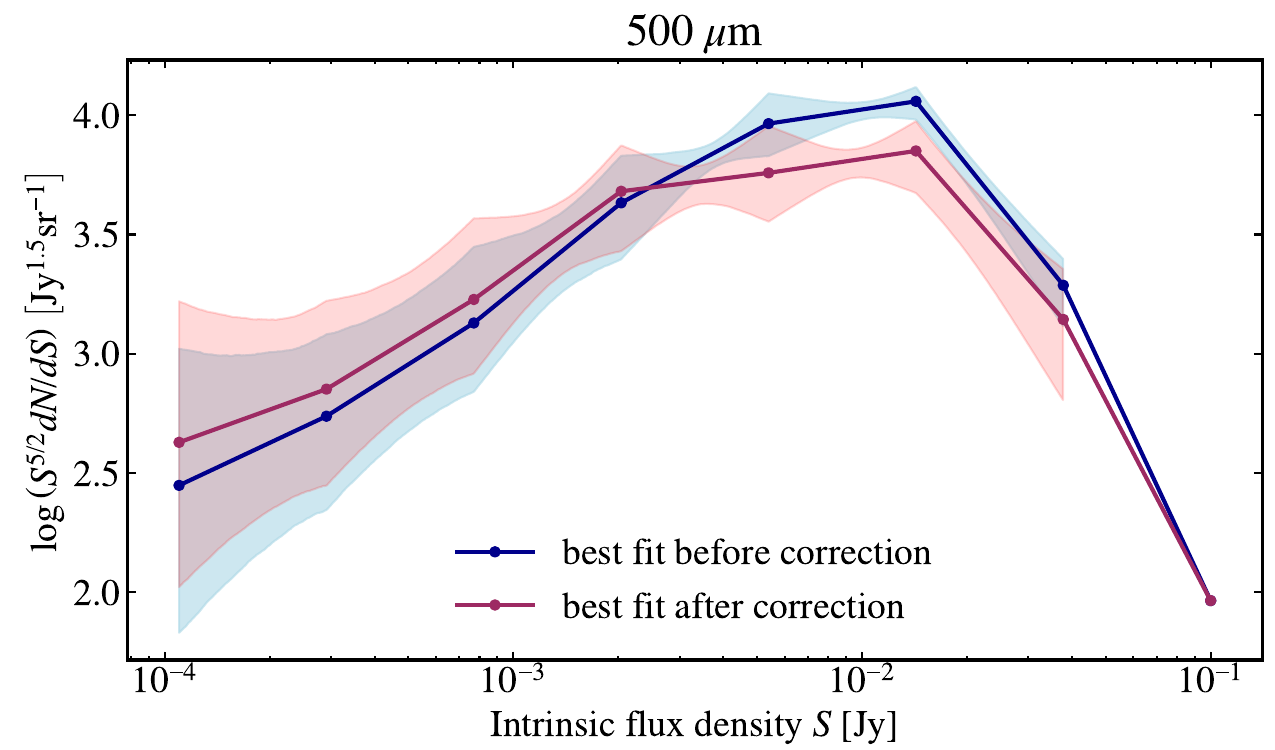}
\caption{Galaxy number count fits before (blue) and after (red) correcting the histograms at 250\,$\mu$m, 350\,$\mu$m and 500\,$\mu$m (top to bottom). Colours and symbols follow Figure~\ref{fig:corr}.}
\label{fig:goodsn_hist}
\end{figure}

We then insert $\theta_0$ values into the relations in Table~\ref{tab:p}, Table~\ref{tab:p_g_250}, and Table~\ref{tab:p_g_350} to determine the parameters in the correction functions (Equation~\ref{eq:corr0}) at 250\,$\mu$m, 350\,$\mu$m, and 500\,$\mu$m. Adding the corresponding correction functions to the original map PDFs and converting to flux histograms, we obtain the clustering-corrected histograms to feed into the $P(D)$ framework. In light of the beam sizes and noise levels at 250\,$\mu$m and at 350\,$\mu$m (different than at 500\,$\mu$m), the adaptive binning and flux cutoffs are carefully chosen following Section~\ref{sec:pdfit}. The error bars in the correction functions are estimated and propagated into the likelihood in the same way as in Section~\ref{sec:pdfit}.

\subsubsection{Models and fits}
\label{sec:model_goodsn}

We choose the node-based model for each band, (see Section~\ref{ch:Models}). Each band has been tested on a simulated map to ensure the efficacy of the model. The brightest node is fixed the simulation when the maximum flux in the map is relatively small, thus the bright galaxy number counts are not well constrained by the map (see Section~\ref{ch:Setup}).

\begin{table*}[htbp!]
\centering
\begin{tabular*}{0.7\textwidth}{rlrlrl}
\hline \multicolumn{2}{c}{$250 \mu \mathrm{m}$} & \multicolumn{2}{c}{$350 \mu \mathrm{m}$} & \multicolumn{2}{c}{$500 \mu \mathrm{m}$} \\
\hline
\renewcommand{\arraystretch}{1.15}
\begin{tabular}{r}
Node \\
{$[\mathrm{mJy}]$}
\end{tabular} & \begin{tabular}{l}
$\log_{10}(\mathrm{d}N/\mathrm{d}S)$ \\
{$\left[\mathrm{sr}^{-1} \mathrm{Jy}^{-1}\right]$}
\end{tabular} & \begin{tabular}{r}
Node \\
{$[\mathrm{mJy}]$}
\end{tabular} & \begin{tabular}{l}
$\log_{10}(\mathrm{d}N/\mathrm{d}S)$ \\
{$\left[\mathrm{sr}^{-1} \mathrm{Jy}^{-1}\right]$}
\end{tabular} & \begin{tabular}{r}
Node \\
{$[\mathrm{mJy}]$}
\end{tabular} & \begin{tabular}{l}
$\log_{10}(\mathrm{d}N/\mathrm{d}S)$ \\
{$\left[\mathrm{sr}^{-1} \mathrm{Jy}^{-1}\right]$}
\end{tabular} \\
\hline 0.01 & $14.75_{-0.88}^{+0.71}$ & 0.07 & $13.06_{-0.83}^{+0.82}$ & 0.11 & $12.53_{-0.61}^{+0.59}$ \\
0.025 & $13.80_{-0.69}^{+0.61}$ & 0.20 & $12.05_{-0.55}^{+0.49}$ & 0.29 & $11.69_{-0.40}^{+0.37}$ \\
0.15 & $12.24_{-0.48}^{+0.50}$ & 0.56 & $11.17_{-0.38}^{+0.42}$ & 0.77 & $11.01_{-0.31}^{+0.34}$ \\
0.49 & $11.36_{-0.35}^{+0.38}$ & 1.57 & $10.51_{-0.22}^{+0.26}$ & 2.04 & $10.41_{-0.25}^{+0.19}$ \\
1.64 & $10.63_{-0.20}^{+0.15}$ & 4.45 & $10.02_{-0.17}^{+0.13}$ & 5.40 & $9.43_{-0.21}^{+0.20}$ \\
5.45 & $9.87_{-0.09}^{+0.10} $ & 12.55 & $8.94_{-0.19}^{+0.16}$ & 14.28 & $8.46_{-0.18}^{+0.13}$ \\
18.13 & $8.80_{-0.18}^{+0.11} $ & 35.42 & $7.58_{-0.26}^{+0.25}$ & 37.80 & $6.70_{-0.34}^{+0.21}$\\
60.21 & $7.00_{-0.45}^{+0.43} $ & 100.00 & $6.17_{-0.74}^{+0.72}$ & & \\
\hline
\end{tabular*}
\caption{Corrected galaxy number count fits on GOODS-N maps at 250\,$\mu$m, 350\,$\mu$m, and 500\,$\mu$m. The lower and upper limits are from the 68\% confidence region (1$\sigma$) of the MCMC fits, except for the first node which is an upper limit due to the assumption of the counts being zero outside the flux range of the fit nodes.}
\label{tab:result}
\end{table*}

We perform the fitting on the corrected GOODS-N histograms using the same \texttt{emcee} configuration as in Section \ref{sec:pdfit}, and show the results of galaxy number count fits before and after correction in Figure~\ref{fig:goodsn_hist} and in Table~\ref{tab:result}. At 10\,mJy, we find the counts are corrected by a factor of $1.12 \pm 0.14$, $1.25\pm 0.38$ and $1.63\pm 0.27$ at 250\,$\mu$m, 350\,$\mu$m, and 500\,$\mu$m. We notice that the clustering bias is indeed the most significant at 500\,$\mu$m, where the beam is the largest.

\section{Discussion}
\label{ch:discussion}
\subsection{Comparison to previous studies}

\begin{figure}[htbp!]
\centering
\includegraphics[width=\textwidth]{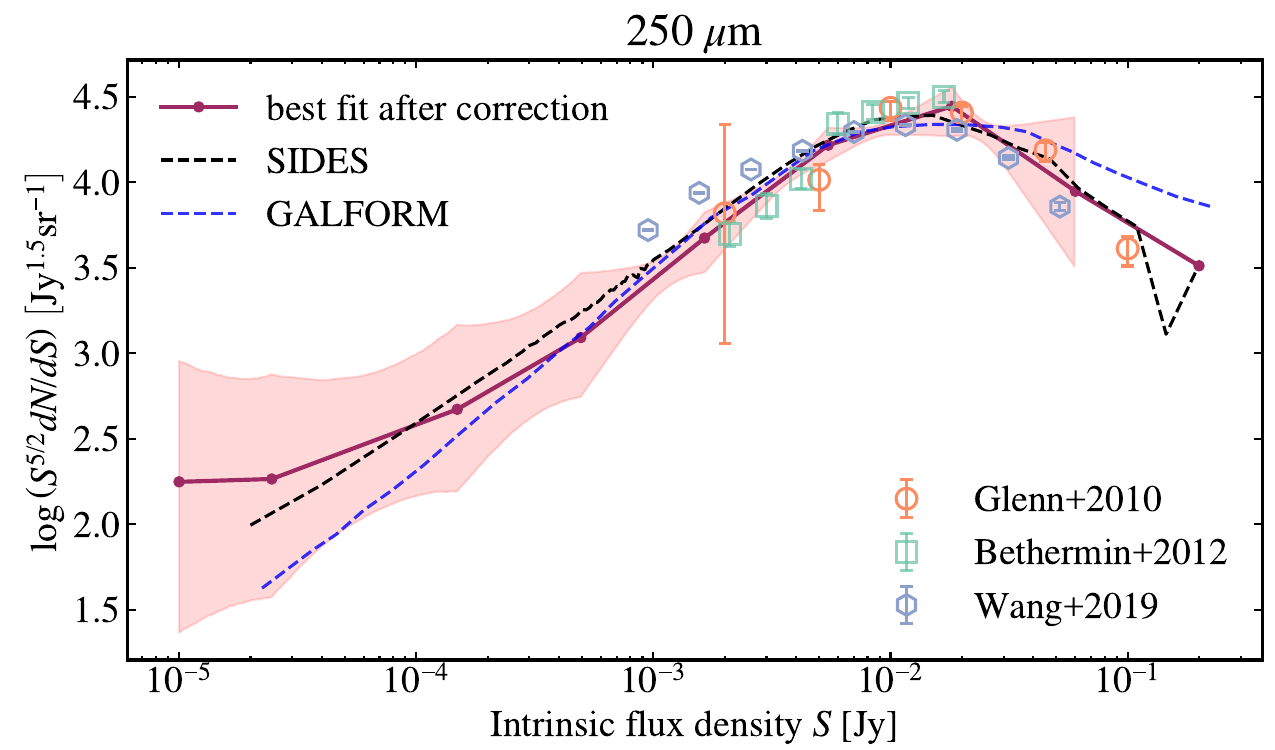}
\includegraphics[width=\textwidth]{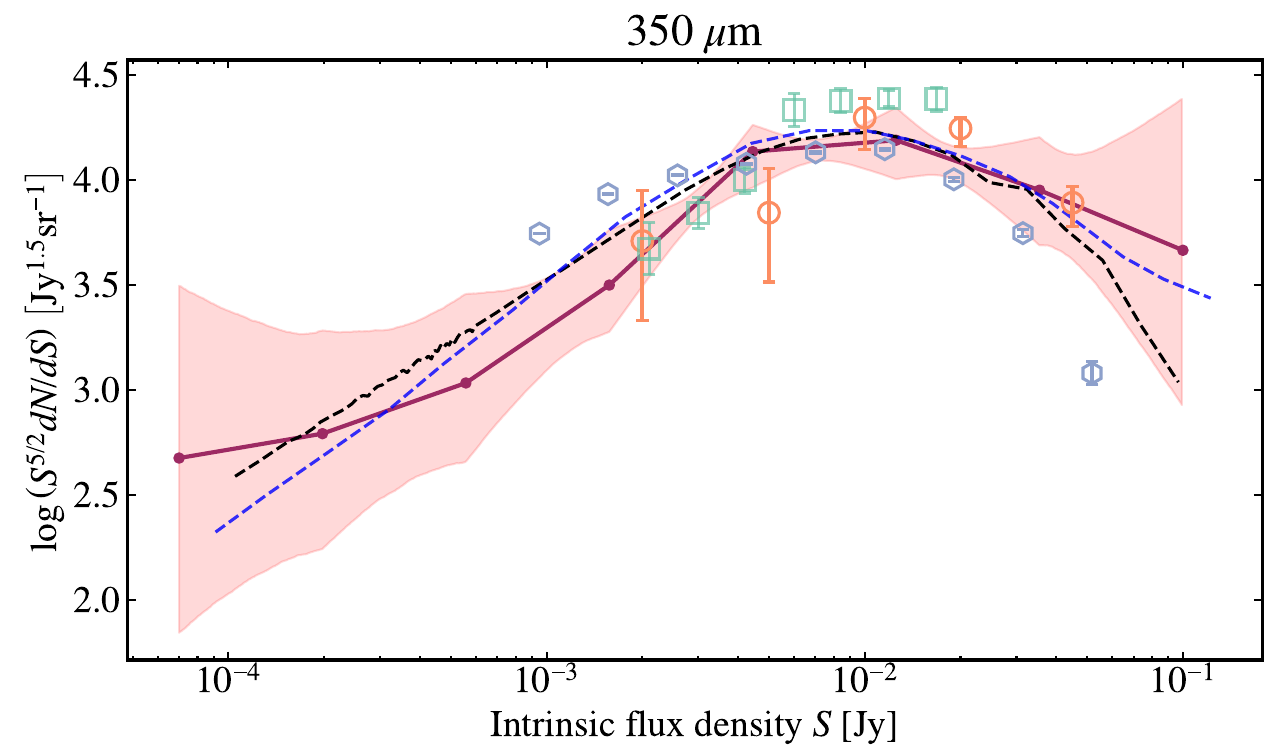}
\includegraphics[width=\textwidth]{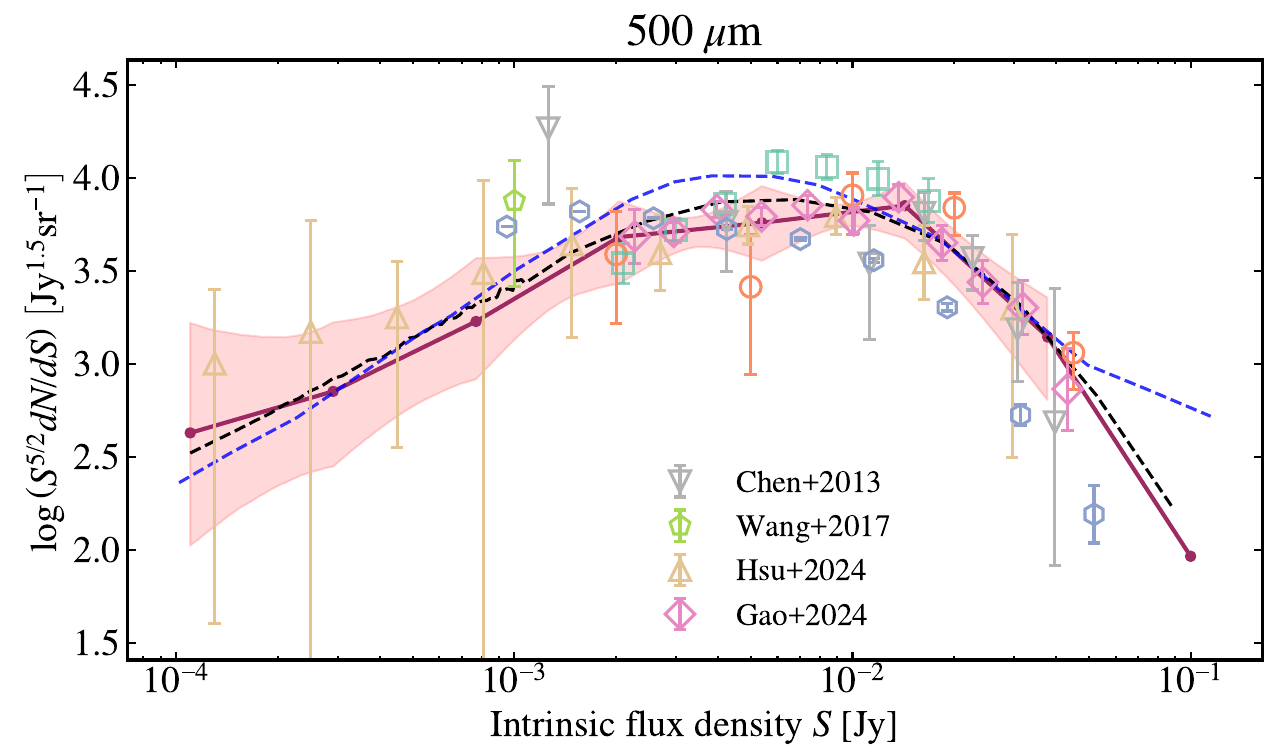}
\caption{Corrected galaxy number counts at 250, 350 and 500\,$\mu$m from top to bottom. The red line and the shaded region are labelled the same as in Figure~\ref{fig:goodsn_hist}. The first node in each panel is an upper limit. Data points are previous measurements at 250, 350, 500\,$\mu$m (\citealt{Glenn2010, Bethermin2012, Wang2019}). For 500\,$\mu$m measurements, deep 450\,$\mu$m counts are also included from \citet{2013ApJ...776..131C}, \citet{Wang2017}, \citet{Hsu2024}, and \citet{gao_scuba-2_2024}.}
\label{fig:goodsn_counts}
\end{figure}

We show the corrected galaxy number counts from this work compared with other measurements at the same wavelength in Figure~\ref{fig:goodsn_counts}. For all three wavebands, we include measurements from \citet{Glenn2010}, \citet{Bethermin2012}, and \citet{Wang2019}. At 500\,$\mu$m, we also compare with deep 450\,$\mu$m measurements using the Submillimetre Common-User Bolometer Array-2 (SCUBA-2; \citealt{2013MNRAS.430.2513H}) camera on the James Clerk Maxwell Telescope (JCMT). We do not convert 450-$\mu$m results to 500\,$\mu$m to avoid assuming a specific shape of the spectral energy distribution for all galaxies.

\citet{Glenn2010} implemented a $P(D)$ fluctuation analysis on three fields of \textit{Herschel}-SPIRE SDP data. GOODS-N is one of the three fields and the deepest one, dominating the measurement below 10\,mJy. We find general agreement with \citet{Glenn2010} in all three bands, the later with no clustering correction included. This is likely due to the noisy data in SDP that contributes to large error bars, as seen in Figure~\ref{fig:goodsn_hist}. Still, the trend of the correction can be seen from the median of the data points, especially at 500\,$\mu$m, where \citet{Glenn2010} lies slightly above our determination at $S\simeq10$\,mJy, but falls below at $S\lesssim10$\, mJy, in agreement with the trend of clustering bias shown in Figure~\ref{fig:goodsn_hist}. This indicates that our clustering correction is effective, while the depth of the full SPIRE data set probes more robustly down to a fainter limit.

\citet{Bethermin2012} used full-depth \textit{Herschel}-SPIRE data and stacked on deep 24\,$\mu$m catalogues in redshift slices. Empirical methods of clustering correction were applied to the stacked flux, based primarily on the clustering of sources in the 24\,$\mu$m catalogue. We find our result in slight tension with the stacking result from \citet{Bethermin2012}, with their counts lying slightly below ours at $S \lesssim 10$\,mJy and above ours at $S \gtrsim 10$\,mJy. This again mirrors the trend of clustering-biased counts in simulations through $P(D)$ analysis, as shown in Figure~\ref{fig:pd}. Although the methods differ in nature and careful examination of their implementations would be needed, we suspect the empirical clustering correction in \citet{Bethermin2012} based on the 24\,$\mu$m catalogue is less accurate for the clustering in the SPIRE bands, which already varies across the three SPIRE bands, as we found in Section~\ref{sec:cf_goodsn}.

At all three SPIRE wavebands, our results are inconsistent with \citet{Wang2019} at both the fainter and brighter ends. At $S \lesssim 10$\,mJy, the counts from \citet{Wang2019} show an excess compared to our work, while a deficit is seen at $S \gtrsim 10$\,mJy. A similar discrepancy is noted in the 500-$\mu$m comparison with 450-$\mu$m results. \citet{Wang2019} performed a multiwavelength deblending based on compiled shorter-wavelength catalogue of the COSMOS field. Because this deblending method inherently relies on positional priors, its accuracy is heavily dependent on the completeness of the shorter-wavelength catalogue, which is known to miss highly obscured submillimetre sources. While a detailed analysis of their algorithm is beyond the scope of this paper, we suggest that this prior-dependency drives the discrepancy. Catalogue incompleteness could result in the simultaneous artificial deficit of bright sources and excess of faint sources seen in the figure. This could be improved with the James Webb Space Telescope (JWST)'s excellent recovery of submillimetre galaxies (e.g. \citealt{Hill2024, 2026MNRAS.545f1961L}).

The 450-$\mu$m results from SCUBA-2 in general show excellent consistency with our 500-$\mu$m results. \citet{Wang2017} presented the preliminary results of the SCUBA-2 Ultra Deep Imaging EAO Survey (STUDIES) compared to \citet{gao_scuba-2_2024}, which  estimated direct counts in the full survey depth. Because of this, we include the full counts from \citet{gao_scuba-2_2024} and only the non-direct counts from \citet{Wang2017}, the latter derived through $P(D)$ fluctuation analysis with a single power-law model without clustering correction. We find that the direct counts from \citet{gao_scuba-2_2024} agree well with our results down to ${\simeq}\,2\,$mJy, while the single power law $P(D)$ results from \citet{Wang2019} are largely consistent with our results at ${\simeq}\,1\,$mJy. \citet{2013ApJ...776..131C} and \citet{Hsu2024} measured combined counts on multiple fields with SCUBA-2, \citet{2013ApJ...776..131C} including two lensing-cluster fields and one blank field, while \citet{Hsu2024} extended this to four lensing-cluster fields and the same blank field. In general, we find excellent agreement with \citep{Hsu2024} down to ${\simeq}\,0.1\,$mJy, and we are consistent with \citep{2013ApJ...776..131C} except for their faintest counts around ${\simeq}\,1.3\,$mJy.
Overall, we find great consistency with the direct, non-direct, and multi-field counts through deep SCUBA-2 observations at 450\,$\mu$m, demonstrating the constraining power of $P(D)$ with the clustering correction proposed in this work.

Finally, we note that at 250\,$\mu$m where the beam is the smallest, our results at the faint end show a trend towards flattening of the Euclidean-normalised counts, which is expected as the counts approach the faint-end turnover. We anticipate that applying this method on deeper observations will fully resolve the extragalactic background light (EBL), and hence provide tighter constraints on models of galaxy formation.

\subsection{Comparison to Simulations}

In Figure~\ref{fig:goodsn_counts}, we also show the counts from SIDES and \textsc{GALFORM} \citep{2019MNRAS.487.3082C}. The SIDES counts show excellent consistency with our results for all three wavebands, despite a slight overestimation around $2\,$mJy at 250 and 350\,$\mu$m and around $4\,$mJy at 500\,$\mu$m. Semi-analytic models (SAMs) such as \textsc{GALFORM} historically struggle to reproduce observed populations of submillimetre galaxies. This tension is largely attributed to the complex treatment of baryonic feedback and dust properties within the simulations \citep[e.g.][]{Lovell2021}. We find that the \textsc{GALFORM} results reproduce the observed counts well at 250\,$\mu$m, while an excess is found at 1\,mJy $\lesssim S\lesssim10$\,mJy for 350\,$\mu$m and 500\,$\mu$m. Resolving this discrepancy will require rigorous scrutiny of the underlying physical prescriptions. Consequently, our results emphasise that submillimetre number counts remain a stringent and highly valuable diagnostic for testing theories of galaxy formation.

Encouragingly, more recent SAMs and hydrodynamic simulations have made progress in reconciling simulated populations of dusty star-forming galaxies with certain observational constraints \citep[e.g.][]{Lovell2021,2025MNRAS.542.2808A}. A direct comparison of our clustering-corrected \textit{Herschel}-SPIRE counts against predictions from these updated simulations will therefore provide interesting constraints on galaxy evolution models in future.

\subsection{Systematic effects}
\label{sec:sys}

The application of $P(D)$ fluctuation analysis requires any variation of the beam to be small and that the noise is nearly Gaussian, which are generally true for {\it Herschel}-SPIRE blank-field observations. \citet{Glenn2010} tested the effect of these assumptions on $P(D)$ analysis and found that a constant Gaussian beam was a good approximation for a small field like GOODS-N. Furthermore, the beam profile in the final SPIRE data is highly stable and well-characterised \citep{Griffin2010}, minimising its systematic impact. Similarly, \citep{Glenn2010} simulated noise variations in SPIRE observations and found it to be negligible for $P(D)$ results.

In this work, we introduced a correction process using the SIDES simulations. As mentioned in Section~\ref{sec: corr}, we incorporated an additional 10\% systematic uncertainty to account for potential clustering effects not captured by the simulation, such as the assumption of a single power-law model for the angular correlation function, $w(\theta)$.

Finally, the roughly 5\% SPIRE calibration uncertainty \citep{Bendo2013, Hopwood2015} has been neglected in our $P(D)$ analysis, since it acts as a global scaling of the flux density and does not meaningfully alter the shape of the pixel histogram. Additionally, the propagated effect of this calibration error on the final number counts is entirely subdominant compared to the existing uncertainties.

\subsection{Field-to-field variance}

Accounting for field-to-field variance is important for measuring galaxy counts due not only to the variation of the counts in different fields, but also the variation of the clustering.  In this work, we address this by measuring the counts in the data themselves, but using an estimate of the clustering scale, $\theta_0$, from averaging over a much larger field.  We tested the sensitivity of the correction function by shifting $\theta_0$ by $\pm 2\,\sigma$ from its average value mapped from the power spectrum (see Section~\ref{sec:cf_goodsn}). We find the variations in the corresponding correction function are smaller than its intrinsic uncertainty (shown in Tables~\ref{tab:p}, \ref{tab:p_g_250}, and \ref{tab:p_g_350}).

\subsection{Gravitational lensing}

Gravitational lensing can visibly alter the observed pixel histogram by boosting faint sources into brighter flux bins.
At the extremely bright end, strong lensing results in a significant overestimation of the bright galaxy counts \citep{2010MNRAS.406.2352L, Paciga2009, 2017MNRAS.465.3558N}. However, this strong lensing regime mostly biases the brighter counts, where the number-count slope is steep \citep[see e.g.][]{2001PhR...340..291B}, dominating above 100\,mJy at 350 and 500\,$\mu$m \citep{Bethermin2017}. This is higher than the brightest source in GOODS-N and significantly above the bright-end histogram cut-off (${\simeq}\,10$\,mJy; see Sections~\ref{sec:pdfit} and \ref{sec:cf_goodsn}) we adopted in our $P(D)$ analysis of the simulated 2\,deg$^2$ maps or the observed GOODS-N maps.

In the faint regime, weak lensing introduces a magnification bias that acts as an extra term for $w(\theta)$ \citep[e.g.][]{2013MNRAS.429.3230H}. The effect of this extra term mimics that of clustering by widening the $P(D)$ distribution. However, the spatial clustering of galaxies dominates over the weak lensing signal in these submillimetre wavelengths by roughly two orders of magnitude \citep[e.g.][]{Wang2011}.

\subsection{More complicated clustering functions}
\label{sec:diss_clustering}

We have demonstrated that our correction method performs well for realistic clustering typical of blank fields ($\theta_0$ up to around 1\,arcsec. Within this range, our proposed functional form of the clustering correction (Equation~\ref{eq:corr0}) well characterises the clustering bias induced in the shape of the $P(D)$ histogram (Figure~\ref{fig:res_fit}). However, we find that the correction function breaks down when clustering becomes even stronger, suggesting that a more accurate correction form would be required for extremely clustered environments, such as galaxy clusters.

Throughout this work, we have adopted a single power-law functional form for the 2PCF, $w(\theta)$, to describe galaxy clustering in blank fields. While this simple model is well-motivated, in practice the clustering strength will depend on galaxy properties, i.e., more massive (or brighter) galaxies tend to cluster more strongly.
Furthermore, when extending to larger scales the simple power-law approximation will break down.
This all suggests that modelling realistic clustering with multi-component power-law forms may yield further improvements to the correction.

\subsection{Other future improvements}
All of the issues discussed in the above subsections are areas where improvements could be made in future. In addition, the shot-noise part of the power spectrum contains some useful information, which we have been neglecting here. We confirmed that the level is in rough agreement with what is expected from our counts, but in future it should be possible to include the shot-noise level in a self-consistent analysis of the 1- and 2-point statistics.  Another direction for future study is the direct constraints on galaxy evolution models, where one could make use of the joint $P(D)$ distribution for the three SPIRE wavebands.

\section{Conclusions}
\label{ch:conclusions}

We have presented a framework for the clustering correction of galaxy number counts derived from $P(D)$ fluctuation analysis, combining 1- and 2-point statistics in the maps to maximise the information gain. Validated against \textit{Herschel}-SPIRE observations using SIDES simulations, the method enables the robust recovery of number counts in the sub-mJy regime, a range otherwise inaccessible due to confusion noise. Applying this framework to data in the GOODS-N field at 250, 350, and 500\,$\mu$m, we find that clustering has a significant impact on number counts, with a correction factor as high as $\sim 1.6$ near 10\,mJy at 500\,$\mu$m. The effect is progressively smaller at 350\,$\mu$m and 250\,$\mu$m, reflecting the decrease in beam size. Our clustering-corrected counts are in excellent agreement with deep SCUBA-2 observations at 450\,$\mu$m. This method is broadly applicable to observations spanning radio to far-infrared wavelengths, provided the beam and noise properties are well characterised, making it a valuable tool for maximising the science return of confusion-limited surveys, including facilities such as {\it Spitzer}, SCUBA-2, the upcoming Cerro Chajnantor Atacama Telescope, as well as new radio surveys.

\begin{acknowledgement}
This paper is dedicated to our colleague Jasper Wall, who was involved in early discussions of this work. 
We thank George Wang for fruitful discussions and helpful comments on the manuscript and figures.
The {\it Herschel\/} spacecraft was designed, built, tested, and launched under a contract to ESA managed by the Herschel/Planck Project team by an industrial consortium under the overall responsibility of the prime contractor Thales Alenia Space (Cannes), and including Astrium (Friedrichshafen) responsible for the payload module and for system testing at spacecraft level, Thales Alenia Space (Turin) responsible for the service module, and Astrium (Toulouse) responsible for the telescope, with in excess of a hundred subcontractors.
SPIRE was developed by a consortium of institutes led by Cardiff University (UK) and including Univ.\ Lethbridge (Canada); NAOC (China); CEA, LAM (France); IFSI, Univ. Padua (Italy); IAC (Spain); Stockholm Observatory (Sweden); Imperial College London, RAL, UCL-MSSL, UKATC, Univ.\ Sussex (UK); and Caltech, JPL, NHSC, Univ. Colorado (USA). This development has been supported by national funding agencies: CSA (Canada); NAOC (China); CEA, CNES, CNRS (France); ASI (Italy); MCINN (Spain); SNSB (Sweden); STFC, UKSA (UK); and NASA (USA).
We acknowledge the use of the SIDES (Simulated Infrared Dusty Extragalactic Sky) simulation.
\end{acknowledgement}

\paragraph{Funding Statement}

This research was supported by the Natural Sciences and Engineering Research Council of Canada.

\paragraph{Competing Interests}

None.

\paragraph{Data Availability Statement}

The \textit{Herschel}-SPIRE observations of the GOODS-N field are publicly available in HELP at \url{https://hedam.lam.fr/HELP/}. The simulated mock catalogues were obtained from the publicly available SIDES (Simulated Infrared Dusty Extragalactic Sky) simulation framework \citep{Bethermin2017} at \url{https://gitlab.lam.fr/mbethermin/sides-public-release}. The custom code used to perform the $P(D)$ fluctuation analysis is available from the corresponding author upon reasonable request.


\bibliography{correct}

\appendix

\end{document}